\documentclass[fleqn,usenatbib]{mnras}

\usepackage{newtxtext,newtxmath}

\usepackage[T1]{fontenc}

\DeclareRobustCommand{\VAN}[3]{#2}
\let\VANthebibliography\thebibliography
\def\thebibliography{\DeclareRobustCommand{\VAN}[3]{##3}\VANthebibliography}

\usepackage{graphicx}	
\usepackage{amsmath}	

\usepackage{subfigure}
\usepackage{booktabs}
\usepackage{multirow}
\usepackage[flushleft]{threeparttable}

\newcommand{\stagger}{\texttt{Stagger}}
\newcommand{\garstec}{\texttt{GARSTEC}}

\newcommand{\mean}[1]{\ensuremath{\langle #1 \rangle}}

\title[Coupling 1D evolution with 3D simulations]{Coupling 1D stellar evolution with 3D-hydrodynamical simulations on-the-fly III: stellar evolution at different metallicities}

\author[Y.~Zhou et al.]{
Yixiao Zhou,$^{1,2}$\thanks{E-mail: yixiao.zhou@qq.com}
Jakob L. R{\o}rsted,$^{2,3}$\thanks{Formerly Jakob R{\o}rsted Mosumgaard}
Achim Weiss,$^{4}$
Andreas C. S. J{\o}rgensen,$^{5,6}$
Cis Lagae,$^{7,8,9}$
\newauthor
Luisa F. Rodr{\'i}guez D{\'i}az,$^{2}$
Yaguang Li,$^{10}$
Mark L. Winther,$^{2}$
Jens R. Larsen,$^{2}$
J{\o}rgen Christensen-Dalsgaard,$^{2}$
\newauthor
Oleg Kochukhov,$^{11}$
Karen R.\ Pollard,$^{1}$
Tanda Li$^{12,13}$
\\
$^{1}$School of Physical and Chemical Sciences --- Te Kura Mat{\=u}, University of Canterbury, Private Bag 4800, Christchurch 8140, Aotearoa, New Zealand\\
$^{2}$Stellar Astrophysics Centre, Department of Physics and Astronomy, Aarhus University, Ny Munkegade 120, DK-8000 Aarhus C, Denmark\\
$^{3}$Aarhus Astronomy Data Centre (AADC), Department of Physics and Astronomy, Aarhus University, Denmark\\
$^{4}$Max Planck Institute for Astrophysics, Karl-Schwarzschild-Str. 1, D-85741 Garching, Germany\\
$^{5}$Department of Mathematics, Imperial College London, Exhibition Rd, South Kensington, London SW7 2BX, UK\\
$^{6}$I-X Centre for AI In Science, Imperial College London, White City Campus, 84 Wood Lane, London W12 0BZ, UK\\
$^{7}$Department of Astronomy, Stockholm University Albanova University Center, 106 91 Stockholm, Sweden\\
$^{8}$Department of Physics, University of Warwick, Coventry CV47AL, UK\\
$^{9}$Center for Exoplanets and Habitability, University of Warwick, Coventry CV47AL, UK\\
$^{10}$Institute for Astronomy, University of Hawaii, 2680 Woodlawn Drive, Honolulu, HI 96822, USA\\
$^{11}$Department of Physics and Astronomy, Uppsala University, Box 516, 751 20 Uppsala, Sweden\\
$^{12}$School of Physics and Astronomy, Beijing Normal University, Beijing, 100875, China\\
$^{13}$Institute for Frontiers in Astronomy and Astrophysics, Beijing Normal University, Beijing 102206, China
}

\date{Accepted XXX. Received YYY; in original form ZZZ}

\pubyear{\the\year{}}

\begin{document}
\label{firstpage}
\pagerange{\pageref{firstpage}--\pageref{lastpage}}
\maketitle

\begin{abstract}
  A major weakness in one-dimensional (1D) stellar structure and evolution modeling is the simplified treatment of convection, which leads to erroneous near-surface stratification and considerable uncertainties in predicted effective temperatures and luminosities of low-mass stars. 
  In a series of preceding works, a novel method for coupling 1D stellar structural models with a grid of 3D surface convection simulations during stellar evolution was developed, at solar metallicity. This 1D-3D coupling method slightly shifts evolutionary tracks relative to standard calculations, meanwhile providing oscillation frequencies that agree more closely with asteroseismic observations. 
  Here we extend this method to model metal-poor and metal-rich FGK-type stars, by implementing interpolations on-the-fly across metallicity ($\rm -3 < [Fe/H] < 0.5$) for mean 3D models during stellar evolution. We demonstrate quantitatively that the fundamental stellar parameters modeled within our framework are insensitive to the mixing-length parameter. A 20\% change in the mixing-length parameter results in evolutionary tracks with a temperature shift of less than 30 K, compared to a difference of over 200 K in standard evolution calculations. Our extension is validated against eclipsing binary systems with extremely precise observational constraints as well as stars in binaries with asteroseismic data. Using a fixed mixing-length parameter that merely governs convective heat transport in the near-adiabatic layers, the 1D-3D coupling method successfully reproduces most observational constraints for all target stars. 
  Coupling 1D stellar evolution models with 3D simulations greatly reduces uncertainties associated with the choice of atmosphere boundary conditions and mixing-length parameters, hence offering a powerful tool for characterizing stars with seismic measurements and determining ages for globular clusters.
\end{abstract}




\section{Introduction}

  1D models of stellar structure and evolution are among the most important instruments in astrophysics. They are capable of predicting the production and transport of chemical elements \citep{2005ARA&A..43..435H,1997ARA&A..35..557P}, pulsation frequencies \citep{2010aste.book.....A}, solar neutrino fluxes \citep{1988RvMP...60..297B}, among others. Moreover, they are a major method for determining the age of stars, which is a crucial quantity for revealing the evolution history of our Galaxy \citep{2018MNRAS.475.5487S,2020NatAs...4..382C} as well as population analysis for star clusters \citep{2018MNRAS.473.2688M}. 
  Nevertheless, the predicting power of stellar evolution calculations is limited by several defects, with the modeling of the near-surface convective region being the most notable one for low-mass stars. Most stellar evolution codes adopt the mixing-length theory (MLT, \citealt{1958ZA.....46..108B}) to describe stellar convection. As a phenomenological theory, the MLT pictures heat transfer in the convective region as discrete ``fluid parcels'' that travel vertically with a distance determined locally, called the mixing length, before merging into their surroundings. The MLT performs well in the deep convection region where the temperature gradient is very close to adiabatic, but it is well-known to fail near the stellar surface where heat transport is far from adiabatic and asymmetries between upflows and downflows are prominent. For example, stellar structure models of solar-type and giant stars computed with MLT predict systematically higher pressure-mode oscillation frequencies than observations at high frequencies, referred to as the asteroseismic surface effect \citep{1996Sci...272.1286C}. This mismatch partly originates from incorrect pressure and density stratification near the stellar surface. Moreover, the convective velocity given by MLT abruptly drops to zero at boundaries between convective and radiative zones, leading to unphysical hard boundaries without convective overshoot. Additionally, MLT contains several free parameters including the mixing-length parameter $\alpha_{\rm MLT}$, which is one of the most uncertain factors in stellar evolution calculations \citep{2023Galax..11...75J}.
  Another associated uncertainty in stellar evolution calculations is the choice of surface boundary condition. Physical quantities at the outermost point of the stellar structural model are determined by the stellar model atmosphere. \citet{2018ApJ...860..131C} found that different atmosphere boundary conditions could lead to $\sim 100$ K effective temperature offset at the red-giant branch of stellar evolution. This adds further complications to the interpretation of observational data for red-giant stars.
  
  Considerable efforts have been made to obtain a more realistic description of convection, by numerically solving the equations of hydrodynamics and radiative transfer in 3D space and time (see \citealt{2009LRSP....6....2N} and references therein). 3D numerical simulations reveal that convection operates in a fundamentally different way than what is suggested by MLT -- the stellar envelope convection is driven by cooling at the surface, which generates finger-like downflows that merge as they descend into deeper convective regions (e.g., \citealt{2017ApJ...845L..23K}). These downward plumes move faster but occupy less space relative to upflows, thus resulting in a downward net kinetic energy flux.
  Solar surface convection simulations have been tested extensively against observations and proven to be superior to their 1D counterpart in all aspects. Without the need for tunable parameters necessary in 1D line formation calculations, \citet{2000A&A...359..729A} reported an excellent match when comparing the detailed spectral lines predicted from 3D solar simulations with observations. \citet{2013A&A...554A.118P} carried out a detailed validation of 3D hydrodynamical models of the solar lower atmosphere and demonstrated that the center-to-limb variations predicted by the 3D solar model agree remarkably well with corresponding observations, indicating the modeled temperature stratification near the optical depth unity is highly realistic.
  Motivated by the success of the solar case, grids of 3D surface convection simulations such as the \texttt{CIFIST} grid \citep{2009MmSAI..80..711L}, the \citet{2013ApJ...769...18T} grid, and the \stagger{} grid \citep{2013A&A...557A..26M,2024A&A...688A.212R} have been computed, covering the main-sequence and giant phase of low-mass stars. These 3D grids have been applied to diverse topics in the physics of low-mass stars across the Hertzsprung-Russell (HR) diagram, including the analysis of chemical composition \citep{2019A&A...630A.104A,2024MNRAS.528.5394W}, granulation and oscillation properties \citep{2022MNRAS.514.1741R,2020MNRAS.495.4904Z}.
  
  Despite the weakness in describing turbulent convection near the stellar surface, 1D  calculations of stellar evolution will not be replaced by full 3D hydrodynamics simulations in the near future because the latter are computationally not affordable in evolutionary time scales \citep{2017LRCA....3....1K}. On the other hand, although realistic theories of turbulent convection have been developed by different groups, such as the non-local mixing-length model of \citet{1977ApJ...214..196G}, the non-local time-dependent theory of \citet{2021FrASS...7...95X} and \citet{2012ApJ...756...37L} derived from hydrodynamic equations and turbulence theory, those theories are not implemented in most stellar evolution codes due to their complexity. Moreover, advanced theories of turbulent convection involve more equations for various interconnected physical quantities, making it much more difficult to obtain a converged solution.
  
  To this end, utilizing pre-computed 3D surface convection simulations to improve 1D evolution models is arguably a promising way forward. A pioneering effort in this direction was made by \citet{1999A&A...346..111L}, who calibrated mixing-length parameters at various effective temperatures $T_{\rm eff}$, surface gravity $\log g$ and chemical compositions based on the entropy profile of their 2D convection simulations. Similar $\alpha_{\rm MLT}$ calibrations were later presented by \citet{2014MNRAS.445.4366T} and \citet{2015A&A...573A..89M}, based on the \citet{2013ApJ...769...18T} and \stagger{} grid of 3D simulations respectively.
  It is worth noting that mixing-length calibrations based on 3D surface convection simulations all predict that $\alpha_{\rm MLT}$ of cool dwarfs are larger than the solar value, whereas $\alpha_{\rm MLT}$ of warm main-sequence-turn-off stars and red giants is smaller \citep{2014MNRAS.445.4366T,2015A&A...573A..89M}, indicating less efficient convection. A different calibration of $\alpha_{\rm MLT}$ based on stars in binary systems with asteroseismic data by \citet{2018MNRAS.475..981L}, however, concluded that $\alpha_{\rm MLT}$ is more than 10\% larger for red giants than for the Sun. Furthermore, when comparing effective temperatures from the APOGEE-\textit{Kepler} catalog \citep{2018ApJS..239...32P} and theoretical predictions, \citet{2017ApJ...840...17T} discovered a trend between temperature mismatch and [Fe/H] that can only be explained by a positive correlation between $\alpha_{\rm MLT}$ and [Fe/H]\footnote{Unless otherwise specified, ${\rm [A/B]} = \log(n_{\rm A} / n_{\rm B}) - \log(n_{\rm A} / n_{\rm B})_{\odot}$ where $n_{\rm A} / n_{\rm B}$ and $(n_{\rm A} / n_{\rm B})_{\odot}$ represent number density ratio between element A and B in the star and the Sun, respectively.}. This positive correlation was supported by \citet{2018ApJ...858...28V} and \citet{2024ApJ...974...77L} who performed $\alpha_{\rm MLT}$ calibrations across different metallicities based on observational data, but it was not seen in $\alpha_{\rm MLT}$ calibrated from grids of 3D simulations (cf.~Fig.~6 of \citealt{2024ApJ...974...77L}). The underlying reason for tensions between mixing-length parameters calibrated from 3D simulations and observations is not understood.
  
  Instead of applying pre-calibrated $\alpha_{\rm MLT}$ and $T(\tau)$ relations, properties of 3D simulation can be used to inform stellar evolution calculations on-the-fly. 
  \citet{2017MNRAS.472.3264J} developed a robust method to interpolate the horizontal- and time-averaged 3D model atmospheres over the $(T_{\rm eff}, \log g)$ plane. The interpolation scheme therefore enables the coupling of the (interpolated) mean 3D stratification with the 1D interior model at every time step of the stellar evolution. 
  In practice, \citet{2018MNRAS.481L..35J} replaced the near-surface regime of the 1D evolutionary model with the mean 3D model and set the outer boundary condition of the stellar evolution calculation from the mean 3D model at every time step of the stellar evolution. Based on this novel approach of coupling 1D stellar evolution and 3D simulation on-the-fly, \citet[hereafter paper I]{2018MNRAS.481L..35J} obtained a new solar structural model and demonstrated that the outermost layers are consistent with the mean 3D solar atmosphere and the coupled model better reproduces observed solar p-mode frequencies. In the following paper, \citet[paper II]{2020MNRAS.491.1160M} further explored how the 1D-3D coupling method would shift evolutionary tracks and reduce the asteroseismic surface effects with respect to the traditional method. However, both previous works were confined to solar metallicity.
  
  In order to extend the 1D-3D coupling method to the modeling of stars with different metallicities with the aim of determining the age of metal-poor stars which are of particular interest in galactic archaeology, in the present work we implement the interpolation of mean 3D models over chemical composition in our stellar evolution code (Sect.~\ref{sec:modeling}). We compare evolutionary tracks computed based on the 1D-3D coupling method with results of standard evolution calculations in Sect.~\ref{sec:stellar-evolution}. The 1D-3D coupling method and newly implemented metallicity interpolation are validated against stars with different compositions in eclipsing binary systems (Sect.~\ref{sec:validation}).

\section{Modeling} \label{sec:modeling}

\subsection{1D stellar structure and evolution models} \label{sec:1D-model}

  Stellar structure models are computed using the Garching Stellar Evolution Code (\garstec{}, \citealt{2008Ap&SS.316...99W}). \garstec{} is among the first generation of programs for numerical simulation of stellar evolution. It solves the equations of mass, energy conservation, hydrostatic equilibrium, and heat transport together at each (spherically symmetric) shell with the \citet{1965ApJ...142..841H} method (see also \citealt{kippenhahn2012stellar} chapter 12.2). Structural models are evolved by evaluating the variation of element abundances throughout the star with time.
  The relationship between thermodynamic quantities, such as density $\rho$, temperature $T$, and pressure $P$, is described by the equation of state (EOS). \garstec{} is equipped with realistic EOSs for stellar interior calculations, including the OPAL EOS \citep{2002ApJ...576.1064R} and \texttt{FreeEOS}\footnote{Available at \url{http://freeeos.sourceforge.net/documentation.html}}  \citep{2004Irwin...feos1,2012ascl.soft11002I}, which are provided in the form of pre-computed tables for different hydrogen, helium, and heavy element mass fractions. Radiative energy transport is modeled with the diffusion approximation, where the Rosseland mean opacities are supplied from opacity tables tabulated according to their underlying chemical composition. In particular, \garstec{} includes opacity tables with varying $\alpha$ element abundance, with $\rm [\alpha/Fe]$ ranging from -0.2 to 0.6 dex.
  The code has been applied to various topics in stellar physics, such as constructing standard solar models \citep{2017ApJ...835..202V}, detailed modeling of solar-type stars with precise asteroseismic measurements \citep{2015MNRAS.452.2127S}, investigating the helium flash and associated element mixing \citep{2003ApJ...582L..43C}, and nucleosynthesis in asymptotic giant branch stars \citep{2024A&A...687A.260R}. Meanwhile, the code undergoes continuous development. Recent updates include the implementation of a non-local theory of convection \citep{2022A&A...667A..96K,2022A&A...667A..97A,2024A&A...689A.292B} and alternative surface boundary conditions obtained from 3D simulations (\citealt{2018MNRAS.478.5650M}; paper I; paper II).
  
  Throughout this study we use the \texttt{FreeEOS} equation-of-state. Above $\log T = 4.1$ (in K), we use OPAL opacity tables \citep{1996ApJ...464..943I}. At lower temperatures, we gradually switch to the opacity tables of \citet{2005ApJ...623..585F}. However, as will be detailed below, low-temperature opacities rarely enter into stellar evolution calculations due to the location of the outer boundary in our 1D-3D coupling method.
  All evolution calculations are based on the \citet{2009ARA&A..47..481A} metal mixture. For nuclear reaction rates and reaction networks, we adopt default settings in \garstec{} described in \citet{2008Ap&SS.316...99W} Sect.~3.2.1.
  
  Apart from nuclear reactions, the distribution of chemical elements over time is also changed by element diffusion, which consists of gravitational settling (also called pressure diffusion) caused by the pressure gradient; thermal diffusion caused by the temperature gradient; concentration diffusion sources from chemical inhomogeneity; and radiative levitation associated with radiation pressure. In this work, diffusion from the first three mechanisms is considered for H, He, C, N, O, Ne, Mg, Si and Fe. Radiative levitation is not implemented in the code. \garstec{} calculates the diffusion coefficient following the method of \citet{1994ApJ...421..828T}, and evaluates the change of chemical composition due to nuclear burning and element diffusion simultaneously.
  
  Convective energy transport in the stellar interior is treated by means of the MLT in the formulation of \citet{kippenhahn2012stellar}. It is worth noting that the strong superadiabatic region near the stellar surface, where MLT is well known to be problematic, is not modeled by \garstec{} in our approach but supplied from the 3D model. The outer boundary of the stellar interior model is located below the photosphere, at the near-adiabatic convective layers. Pressure and luminosity at the outermost point, as well as their partial derivatives with respect to temperature and radius, are supplied by interpolating mean 3D model atmospheres at every time step of stellar evolution (cf.~Sect.~\ref{sec:3D-model} and \ref{sec:1D-3D}). The choice of outer boundary conditions is the core of our methodology.
  At other convective boundary layers such as the tachocline located at the bottom of the convective envelope, \garstec{} has the option to account for element mixing caused by convective overshoot (or undershoot) that extends beyond the Schwarzschild boundary of convection. The code employs an exponential decay of overshooting diffusion coefficient as suggested in \citet{1996A&A...313..497F}. The overshoot parameter $f_{\rm ov}$ controls the $e$-folding distance of the diffusion coefficient.

\subsection{3D model atmospheres and the interpolation technique} \label{sec:3D-model}

\begin{figure}
\includegraphics[width=\columnwidth]{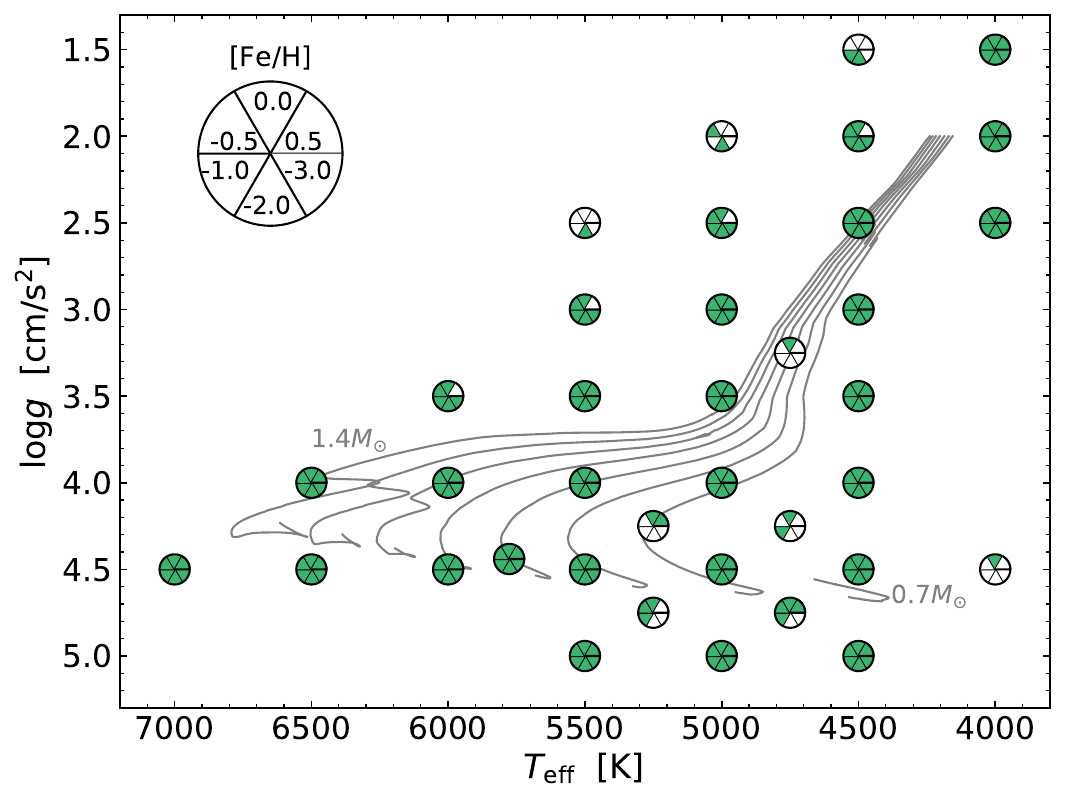}
\caption{Kiel diagram showing the global parameters of the \stagger{}-grid models, with green-shaded portions indicating the 3D model atmospheres used in this work (see also Fig.~5 of \citealt{2024A&A...688A.212R}). Note that the mean effective temperatures of the 3D models are close to, but generally not exact multiples of 500 K.
Evolutionary tracks of 0.7 - $1.4 M_{\odot}$ stars, whose initial chemical compositions are determined from solar calibration (Table \ref{tb:solar-cali}), computed based on the 1D-3D coupling method are depicted in the background. The $0.7 M_{\odot}$ track stops at 20 Gyr.
\label{fig:grid}}
\end{figure}

  The outer boundary conditions of the stellar evolution calculations are provided by the \stagger{}-grid -- a grid of 3D model atmospheres, also called surface convection simulations, of F, G, K-type dwarf and giant stars originally presented by \citet{2013A&A...557A..26M} and subsequently refined and extended by \citet{2024A&A...688A.212R}.
  The grid was generated from the \stagger{} code \citep{1995...Staggercodepaper,2018MNRAS.475.3369C,2024ApJ...970...24S}, a versatile magnetohydrodynamics code that solves the equations of mass, momentum, energy conservation, and the induction equation together with the equation of radiative transfer in three spatial dimensions and time. Magnetic fields were ignored in the calculation of \stagger{}-grid, whereas radiative energy transport was modeled in detail for every mesh and time step. The spatial dependence of the radiative transfer equation was taken into account by solving the equation at different polar and azimuthal angles, while the frequency dependence of the problem was approximated by the opacity binning method (also called multigroup method, see \citealt{2013A&A...557A..26M} Sect.~2 and \citealt{2018MNRAS.475.3369C} Sect.~2.4).
  The \stagger{}-grid was generated based on the \citet{2013ApJ...769...18T} version of the \citet{1988ApJ...331..815M} equation of state. A comprehensive collection of continuous opacity sources, summarized in \citet{2010A&A...517A..49H}, and line opacities from the MARCS opacity sampling data \citep{2008A&A...486..951G,2008PhST..133a4003P} were used in the radiative transfer calculations. All models adopt the \citet{2009ARA&A..47..481A} metal mixture scaled with metallicity. A 0.4 dex enhancement of $\alpha$-element abundances was applied to metal-poor models with $\rm [Fe/H] \leq -1$.
  
  At the current time, the \stagger{}-grid consists of more than 220 3D models with $T_{\rm eff}$ ranges from 3500 to 7000 K, $\log (g / \rm [cm/s^2])$ from 1.5 to 5, and [Fe/H] from $-4$ to $0.5$. Unlike $\log g$ and [Fe/H], which are input parameters of the simulation, the $T_{\rm eff}$ of each simulation snapshot is derived from the radiative flux at the top of the simulation domain. The effective temperature is therefore an emergent property of the simulation fluctuating with time. In this work, we utilize the whole grid except for the $\rm [Fe/H] = -4$ ones and relatively shallow simulations whose lower boundary is located at or near the strong superadiabatic layers (Fig.~\ref{fig:grid}). 
  The simulation domain is a cuboid box with $240 \times 240$ mesh points in the horizontal plane and 230 meshes along the vertical.
  Although the characteristic length could differ by several orders of magnitude across the HR diagram, the horizontal area of the simulation was optimally determined such that at least ten granules are enclosed at any time of each simulation \citep[Sect.~2.1]{2013A&A...557A..26M}. Vertically, 3D model atmospheres cover a small but nevertheless important region of the star. All our selected models span from the near-adiabatic convective layers (typically, Rosseland optical depth $\tau_{\rm Ross} \sim 10^6$ to $ 10^7$ at the bottom boundary) to the base of the chromosphere, with vertical size being at least 15 times the photospheric pressure scale height. The fact that \stagger{}-grid models extend relatively deep below the surface (compared to most 1D model atmospheres) enables us to match 1D and 3D models at the near-adiabatic region, which makes the stellar evolution calculation insensitive to the value of mixing-length parameter (cf.~Sect.~\ref{sec:solar-evol}).
  Each 3D model contains at least 100 simulation snapshots evenly sampled in time. The time sequence of these simulations covers at least 6 periods of the main pressure mode (p-mode) in the simulation box, corresponding to at least $\sim 25$ convective turnover time (defined as pressure scale height divided by convective velocity) at the photosphere. 
  
  We average 3D model atmospheres in space and time in order to apply them to stellar evolution calculations. Here the horizontally averaged model is obtained by taking the mean value at constant geometric depth, i.e., the simple horizontal average. 
  Compared to other averaging methods, such as averaging over constant optical depth, pressures and densities obtained through the simple horizontal and time averaging method very closely obey hydrostatic equilibrium (cf.~\citealt{2013A&A...560A...8M} Fig.~A.2 and \citealt{2023A&A...677A..98Z} Fig.~9). We refer to the simple horizontal- and time-averaged 3D model as mean 3D or \mean{\rm 3D} model hereinafter. The effective temperature of the mean 3D model is the time-averaged value from all simulation snapshots.
  
  Using mean 3D models at every time step of the stellar evolution calculation requires interpolating the model grid at different $T_{\rm eff}$, $\log g$, and metallicities. \citet{2017MNRAS.472.3264J,2019MNRAS.484.5551J} conceived and developed a robust interpolation method to obtain a mean 3D model for any given fundamental stellar parameter covered by the grid. The key component of their interpolation scheme is the ``density inflection region'' (called density jump in \citealt{2017MNRAS.472.3264J}), where the density gradient is relatively small, or, for some stars, a density inversion occurs (e.g., \citealt[Fig.~2]{2007A&A...469..687C}). For F, G and K-type stars, the density inflection is associated with the large superadiabatic temperature gradient at the top of the hydrogen ionization zone just below the stellar surface. Quantitatively, the location of the density inflection corresponds to the local minimum of the partial derivative of density with respective to pressure, $\partial\ln\rho / \partial\ln P$. 
  For each mean 3D model, all quantities are scaled based on their value at the density inflection. The scaled structure $f^{\prime} = f/f_{\rm di}$, where $f_{\rm di}$ could be density, pressure, temperature etc.~at the location of density inflection, is highly similar across the whole \stagger{}-grid (see Fig.~2 of \citealt{2017MNRAS.472.3264J}), thus ensuring a stable and robust interpolation.
  
  In practice, we first trim all \mean{\rm 3D} models by locating the minimum superadiabatic temperature gradient in the optically thick region then discarding all layers beneath it. The reasoning is that the superadiabatic temperature gradient $\nabla_T - \nabla_{\rm ad}$, where $\nabla_T$ and $\nabla_{\rm ad}$ is the actual and adiabatic temperature gradient respectively, should decrease monotonically when moving deeper into the convection zone, hence an increase of $\nabla_T - \nabla_{\rm ad}$ with depth reflects numerical effects from the bottom boundary conditions. At each metallicity, the trimmed \mean{\rm 3D} models define the reference scaled pressures $P^{\prime} = P/P_{\rm di}$ used in the interpolation. To avoid extrapolations, the maximum $P^{\prime}$ is dictated by the trimmed model which has the smallest innermost scaled pressure. For given effective temperature and surface gravity, scaled physical quantities $f^{\prime}$ are interpolated at each value of scaled pressures. The same procedure is repeated for each metallicity value, followed by a 1D interpolation at the target metallicity to obtain the desired scaled stratification. Finally, the interpolated stratification $f^{\prime}(P^{\prime})$ is multiplied by its value at the density inflection $f_{\rm di}$, which is likewise obtained via interpolation, to get the interpolated \mean{\rm 3D} model. 
  
  2D linear interpolation in $(T_{\rm eff}, \log g)$ following the technique described above has been implemented in \garstec{} in papers I and II. Nevertheless, as the \stagger{}-grid has a relatively large interval in effective temperature and surface gravity, the stratification of the interpolated mean 3D model is likely affected by the interpolation method. An optimal interpolation method should give mean 3D structures in best agreement with the ones computed from 3D surface convection simulations. Here we carry out multiple tests at different $(T_{\rm eff}, \log g)$ pairs by taking one \mean{\rm 3D} model out from the grid, constructing an interpolated model at corresponding surface temperature and gravity based on the remaining models, then comparing the interpolated \mean{\rm 3D} model with the actual profile predicted by the 3D simulation. 
  The thus evaluated interpolation errors from the linear and cubic method are shown in Figs.~\ref{fig:itp-diff-solar}-\ref{fig:itp-diff-t47g27m00} in Appendix \ref{app:itp-error} for atmosphere parameters corresponding to the Sun, dwarf and giant stars. Errors below the location of density inflection are the main focus here, since the matching between \mean{\rm 3D} and the stellar interior model takes place in the deeper layers. We find that the mean structure of cool dwarfs is better reproduced by linear interpolation. Conversely, cubic interpolation gives smaller errors for warm dwarfs, whose effective temperature is around and greater than the solar value, as well as for RGB stars. In this study, we opt to employ 2D linear interpolation when $\log g \geq 4$ and 2D cubic interpolation when $\log g \leq 3.5$. A smooth transition\footnote{using a harmonic function with continuous first derivatives} between results from the two interpolation methods is ensured when $3.5 < \log g < 4$ to prevent abrupt changes in the interpolated mean 3D profile that may cause convergence problems in \garstec{}. Therefore, our choice of interpolation method is suitable for cool dwarfs and giants. Meanwhile, we caution that systematic errors due to interpolation are likely larger in the high $T_{\rm eff}$ and $\log g$ regions.
  
  The interpolation in $(T_{\rm eff}, \log g)$ is performed at each metallicity of the grid, which gives an univariate function between metallicity and physical quantities. We base the metallicity interpolation on the metal-to-hydrogen ratio ${\rm [M/H]} = \log_{10} (Z/X) - \log_{10} (Z/X)_{\odot}$ rather than the iron-to hydrogen ratio [Fe/H], where $X, Y, Z$ are the mass fraction of hydrogen, helium and heavier elements, respectively. Interpolation in [M/H] is achieved using the cubic monotonic method of \citet[see also \citealt{2011ApJS..192....3P}]{1990A&A...239..443S}. The thus obtained interpolation results always lie within the range of the input data. The metallicity interpolation is implemented in \garstec{}, making the 1D-3D coupling method applicable to the evolution of stars with different [M/H], and in addition leads to more consistent atmosphere boundary conditions when considering element diffusion in stellar evolution calculations.

\subsection{Coupling 1D stellar evolution with 3D model atmospheres} \label{sec:1D-3D}

\begin{figure}
\centering
\includegraphics[width=0.4\columnwidth]{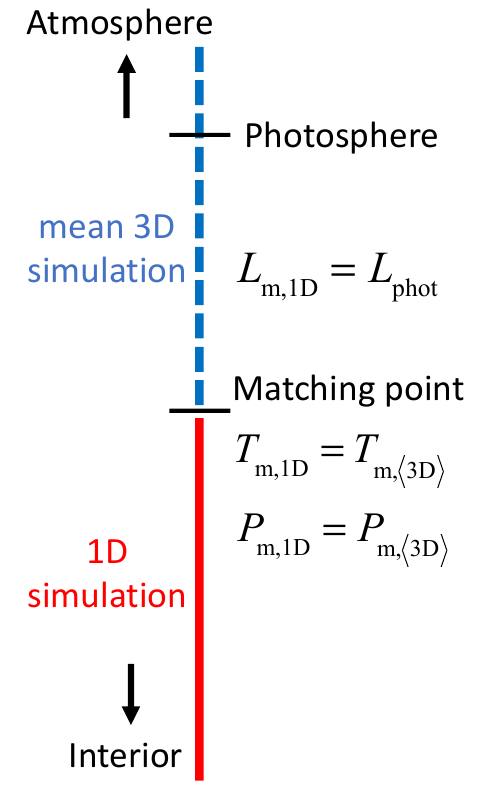}
\caption{Schematic overview of the 1D-3D coupling approach. The mean 3D model is employed as the outer boundary condition for the stellar evolution calculation. The matching point is located well below the photosphere. The subscript ``m'' denotes quantity at the matching point. See also \citet{2019MNRAS.488.3463J} Fig.~1 for a detailed flowchart of the method.
\label{fig:1D3D-overview}}
\end{figure}

  In this section, we summarize how 1D stellar interior models are coupled with \mean{\rm 3D} model atmospheres at every evolution time step. Our adopted approach was originally developed in papers I and II. We refer the readers to these previous works for more details about the methodology and workflow.
  
  First we define the location where the stellar interior and \mean{\rm 3D} atmosphere models are connected, that is, the matching point. As the matching point serves as the outermost shell of the stellar interior model, ideally it should be placed as deep as the 3D model could cover where the temperature profile is closer to adiabatic and the phenomenological treatment of convective heat transport in 1D models is of less concern. In reality, given that (1) numerical effects from the bottom boundary condition have an impact on the deepest layers in 3D simulations, and (2) the interpolation error increases with depth in deep layers of mean 3D models (Figs.~\ref{fig:itp-diff-solar}-\ref{fig:itp-diff-t47g27m00}), an optimal matching point location is therefore determined based on a compromise between these considerations. Here we follow papers I and II to locate the matching point according to pressure at the density inflection. The pressure at the matching point, $P_{\rm m,\mean{3D}}$, is always approximately 15.8 times greater than the pressure at the density inflection, $P_{\rm m,\mean{3D}} = 10^{1.2} P_{\rm di}$.
This criterion ensures the matching between 1D and \mean{\rm 3D} models takes place  
well beneath the optical surface (at $\tau_{\rm Ross} \gtrsim 10^4$) where the superadiabatic temperature gradient $\nabla_T - \nabla_{\rm ad}$ is an order of magnitude smaller than the peak value just below stellar photosphere (Fig.~\ref{fig:sad}). The effect of the matching point location on stellar evolution from the 1D-3D coupling method was thoroughly investigated in paper II.

  It is worth noting that the turbulent pressure, which originates from velocity fluctuations, contributes to the total pressure. The effects of turbulent pressure on the evolution of a one solar-mass star and solar oscillation frequencies were studied by \citet{2019MNRAS.488.3463J}, who concluded that including turbulent pressure has negligible impact on the evolutionary track. Indeed, for G- and K-type stars, turbulent pressure consists less than 5\% of the total pressure below the matching point as predicted by \stagger{}-grid models (e.g.~\citealt[Fig.~21]{2013A&A...557A..26M}), due to smaller velocity fluctuations in deeper layers. The fraction of turbulent pressure increases in warmer stars, but the ratio between turbulent and total pressure is less than 10\% even for the warmest \stagger{}-grid models ($T_{\rm eff} \approx 7000$ K). 
  Although it could be beneficial to include turbulent pressure in the modeling of early F-type stars and their oscillation frequencies, it is not taken into account in stellar interior modeling in the present work. For consistent couplings between 1D and \mean{\rm 3D} models, the turbulent pressure component in \mean{\rm 3D} models is also excluded entirely during the interpolation. We therefore do not distinguish between thermal and total pressure and simply symbolize them as $P$.
  
  Global stellar parameters $T_{\rm eff}$, $\log g$ and [M/H] play an important role in the interpolation of mean \stagger{}-grid models and specifying outer boundary conditions for \garstec{}. In each iteration, surface gravity is evaluated at the matching point. We note that the distance between the matching point and photosphere is less than 2.5\% of the stellar radii (less than 0.5\% of the total radii for dwarf stars) for all models except for the $\log g = 1.5$ ones. Equalizing gravity at the matching point to the surface value is therefore a reasonable approximation and meanwhile consistent with the fact that all atmosphere models were constructed assuming constant gravitational acceleration in the simulation domain.
  Surface metallicity [M/H] is also determined at the matching point, which is justified due to the surface convection zone being well mixed and hence having a uniform chemical composition.
  
  Based on the temperature at the outer boundary of the interior model $T_{\rm m,1D}$, $\log g$, and [M/H], we interpolate the bivariate function $T_{\rm eff}( T_{\rm m,1D}, \log g )$, which is obtained from the relationship between effective temperature and temperature of the matching location for \stagger{} models, followed by the metallicity interpolation to obtain the effective temperature. The thus obtained $T_{\rm eff}$ is then used to derive the mean 3D model as outlined in Sect.~\ref{sec:3D-model}. Nevertheless, the interpolated temperature at the matching point, $T_{\rm m,\mean{3D}}$, might differ from $T_{\rm m,1D}$ because of interpolation errors. That is, the interpolation is not reversible. In view of this, we adjust the effective temperature iteratively until the interpolated temperature at the matching point agrees with $T_{\rm m,1D}$ within 0.01\%.
  The interpolated temperature and (thermal) pressure profile, which extends to the lower stellar atmosphere, is subsequently appended on top of the 1D model. The density stratification above the matching point is evaluated from \texttt{FreeEOS} in \garstec{} based on temperature and pressure of the mean 3D model, rather than directly taken from the mean 3D model, to avoid discontinuities.
  Having established the density and pressure structure, we integrate the equations of mass conservation and hydrostatic equilibrium outwards to calculate stellar radius and enclosed mass above the matching point. The complete stellar structure model yields total luminosity, $L_{\rm phot}$, from the Stefan-Boltzmann law, where the photospheric radius $R$ is the location where $T = T_{\rm eff}$. 
  The aforementioned procedure provides two constraints on the stellar evolution calculation. First, pressure at the outermost shell of the interior model must equal to the \mean{\rm 3D} value at the matching point, $P_{\rm m,1D} = P_{\rm m,\mean{3D}}$. Second, since no nuclear reaction takes place at the near-surface region, luminosity at the top of the 1D model must equal to the photospheric value, $L_{\rm m,1D} = L_{\rm phot}$. The numerical solver in \garstec{} will iteratively alter the interior structure to fulfill the two outer boundary conditions before advancing to the next evolutionary time step. Therefore, our 1D-3D coupling method ensures continuous pressure and temperature at the matching point, as well as a consistent luminosity between the 1D interior model and the full (patched) model throughout the evolution calculation.
  
  In short, as outlined in Fig.~\ref{fig:1D3D-overview}, the 1D-3D coupling method replaces the near-surface regime of the 1D evolutionary model with the (interpolated) mean 3D model, and sets the outer boundary condition of the stellar evolution calculation from the mean 3D model in every time step of stellar evolution.

\section{Stellar evolution and structure} \label{sec:stellar-evolution}

\subsection{Solar calibration and evolution} \label{sec:solar-evol}

\begin{table*}
\centering
\caption{Solar calibration results based on different descriptions of the outer boundary, including the Eddington gray atmosphere, the empirical $T(\tau)$ relation by \citet{1966ApJ...145..174K} and the $T(\tau)$ relation fitted to the model C of \citet[VAL-C]{1981ApJS...45..635V}. Key input parameters of evolution calculations obtained from solar calibrations are listed in the lower part.
\label{tb:solar-cali}}
\begin{threeparttable}
{\begin{tabular*}{0.8\textwidth}{@{\extracolsep{\fill}}lccccc}
\toprule[2pt]
					& Reference	& 1D-3D coupling	& Gray atmosphere	& Krishna Swamy (KS)	& VAL-C
  \\
\midrule[1pt]
  $T_{\rm eff}$ (K)	& 5772 (a)	& 5772			& 5772				& 5772					& 5772
  \\
  $R$ ($10^{10}$ cm)	& 6.957 (a)	& 6.957			& 6.957				& 6.957					& 6.957
  \\
  $X_{\rm surf}$   	& 0.7381 (b)	& 0.7490			& 0.7489				& 0.7489					& 0.7489
  \\
  $Y_{\rm surf}$  	& 0.2485 (c)	& 0.2374			& 0.2375				& 0.2375					& 0.2375
  \\
  $Z_{\rm surf}$  	& 0.0134 (b)	& 0.0136			& 0.0136				& 0.0136					& 0.0136
  \\
  $(Z/X)_{\rm surf}$  & 0.0181 (b)	& 0.01815			& 0.01815				& 0.01815					& 0.01815
  \\
\midrule[1pt]
  $\alpha_{\rm MLT,\odot}$	& ---		& 2.77			& 1.77				& 2.07					& 1.98
  \\
  $X_{\rm init}$ 	& ---		& 0.7232			& 0.7232				& 0.7232					& 0.7232
  \\
  $Y_{\rm init}$  	& ---		& 0.2622			& 0.2622				& 0.2622					& 0.2622
  \\
  $Z_{\rm init}$   	& ---		& 0.0146			& 0.0146				& 0.0146					& 0.0146
  \\
\bottomrule[2pt]
\end{tabular*}}

    \begin{tablenotes}
      \item (a) \citet{2016AJ....152...41P}; (b) \citet{2009ARA&A..47..481A}; (c) \citet{2004ApJ...606L..85B}
    \end{tablenotes}
    
\end{threeparttable}
\end{table*}

\begin{figure*}
\centering
\includegraphics[width=0.49\textwidth]{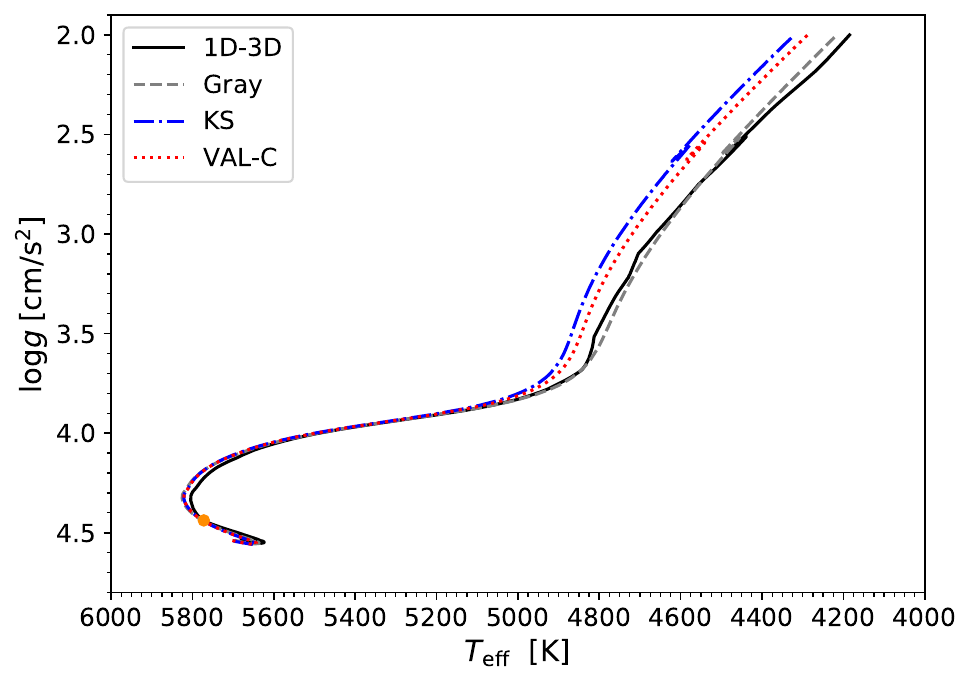}
\includegraphics[width=0.49\textwidth]{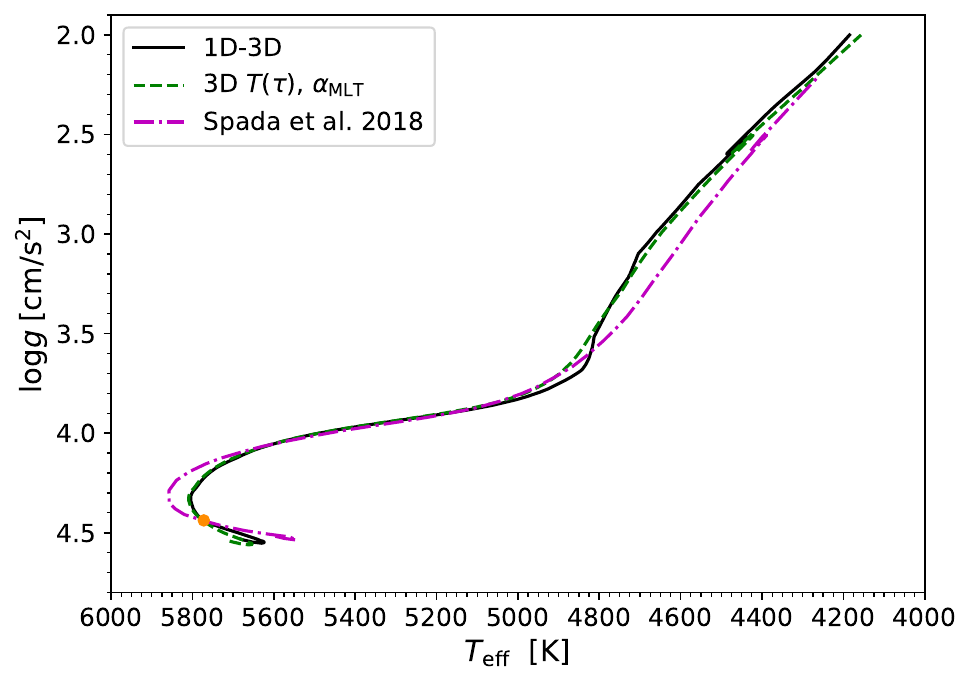}
\caption{Evolution of $1 M_{\odot}$ stars computed with different treatments of the outer boundary and otherwise identical input physics (see Sect.~\ref{sec:solar-evol} for the abbreviations used for the outer boundary conditions). The dash-dotted magenta line in the \textit{right panel} is an exception, representing a model computed with the entropy calibration method adopted from \citet[model ESM1 in their Fig.~10]{2018ApJ...869..135S}. 
For all four tracks in the \textit{left panel}, the mixing-length parameter and initial mass fractions are determined via solar calibration. In the case where the evolution is carried out with 3D-calibrated $T(\tau)$ relations and mixing-length parameters (green dashed line in the \textit{right panel}), the solar calibration sets the scaling factor (cf.~\citealt{2018MNRAS.478.5650M} Eq.~2) while relative values of $\alpha_{\rm MLT}$ follow those determined by \citet{2015A&A...573A..89M}. All tracks pass through the solar parameters (marked with an orange circle) at the present solar age. 
\label{fig:solar-evol}}
\end{figure*}

\begin{figure*}
\centering
\includegraphics[width=0.8\textwidth]{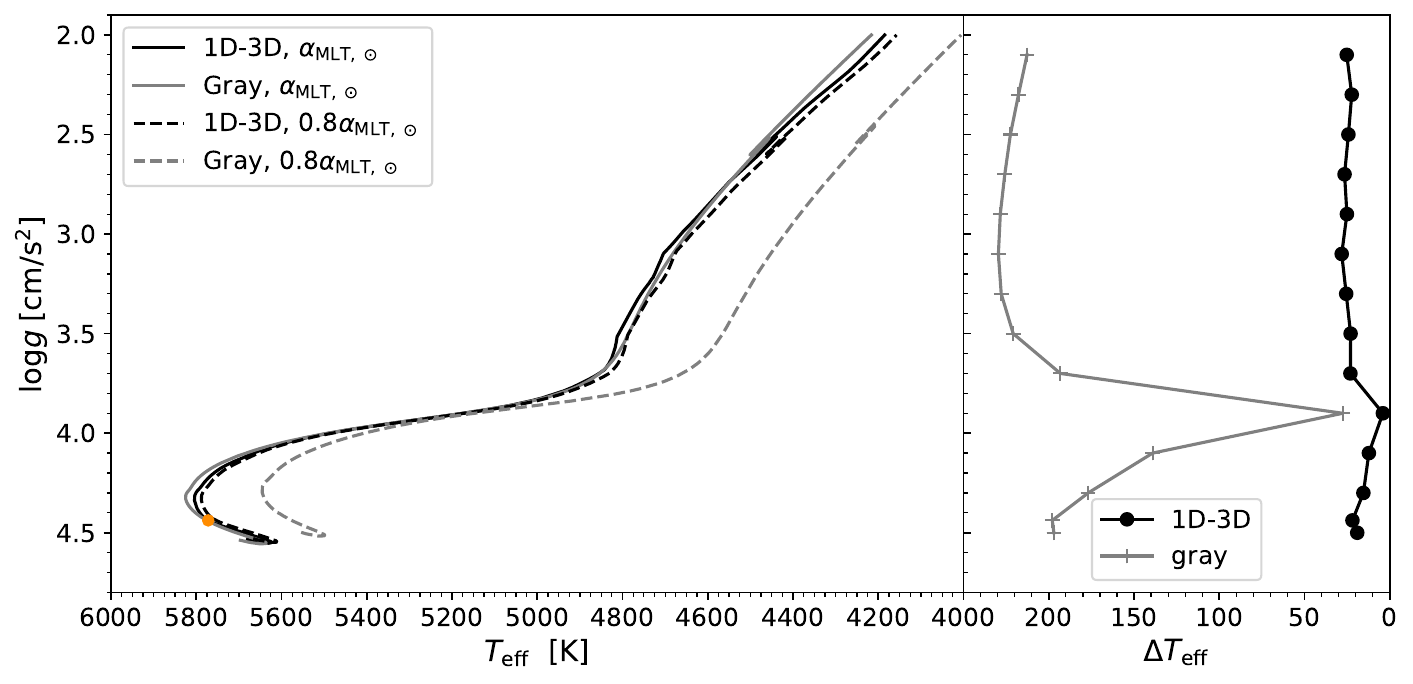}
\caption{The effect of changing $\alpha_{\rm MLT}$ on surface properties of stars for the 1D-3D coupling approach and standard stellar evolution with gray atmosphere. Black and gray solid lines in the \textit{left panel} are evolutionary tracks of solar mass star calculated with solar-calibrated $\alpha_{\rm MLT}$, which are also shown in Fig.~\ref{fig:solar-evol}. The sole difference between black (also gray) dashed and solid line is the adopted $\alpha_{\rm MLT}$ value. 
The absolute difference in $T_{\rm eff}$ at given surface gravity caused by varying $\alpha_{\rm MLT}$ is demonstrated in the \textit{right panel}. In standard evolution calculations, a 20\% relative difference in $\alpha_{\rm MLT}$ shifts $T_{\rm eff}$ by about 200 K on the main-sequence and around 225 K along the giant branch. However, changing $\alpha_{\rm MLT}$ by the same relative amount results in less than 30 K difference in $T_{\rm eff}$ when the 1D-3D coupling method is used.
\label{fig:alphaMLT}}
\end{figure*}

\begin{figure}
\centering
\includegraphics[width=\columnwidth]{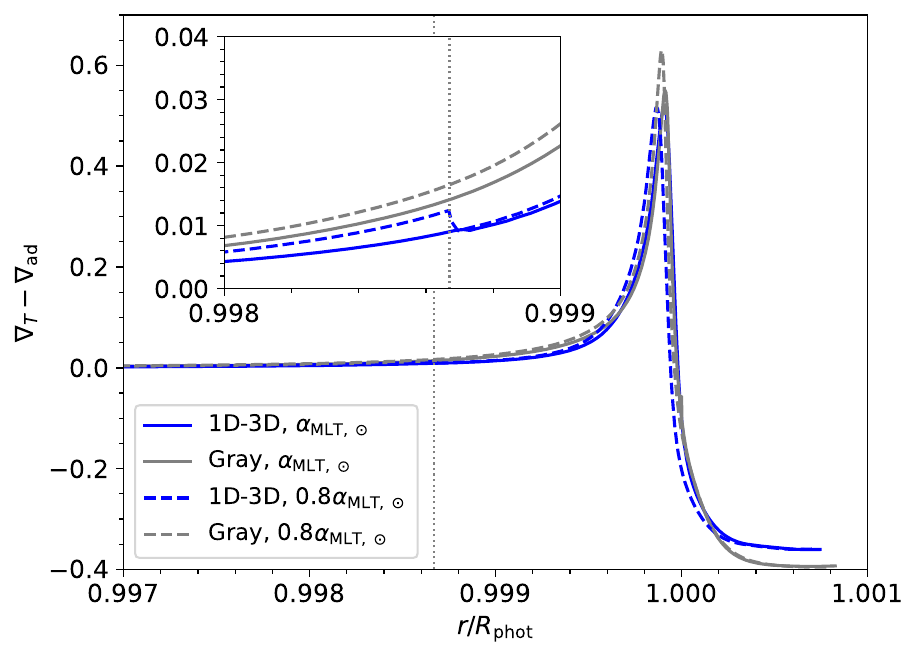}
\caption{The distribution of superadiabatic temperature gradient in the near-surface region, as predicted by \garstec{} solar models constructed with different outer boundary conditions (blue and gray solid line). Corresponding results from models at solar $\log g$ (hence solar radius) but calculated with a smaller $\alpha_{\rm MLT}$ (see dashed lines in Fig.~\ref{fig:alphaMLT}) are shown in dashed lines. The vertical dotted line marks the location of the matching point within the 1D-3D coupling approach, above which the superadiabatic temperature gradient is obtained from the interpolated \mean{\rm 3D} models. The superadiabatic temperature gradient peaks just below the photosphere.
\label{fig:sad}}
\end{figure}

\begin{figure}
\centering
\includegraphics[width=\columnwidth]{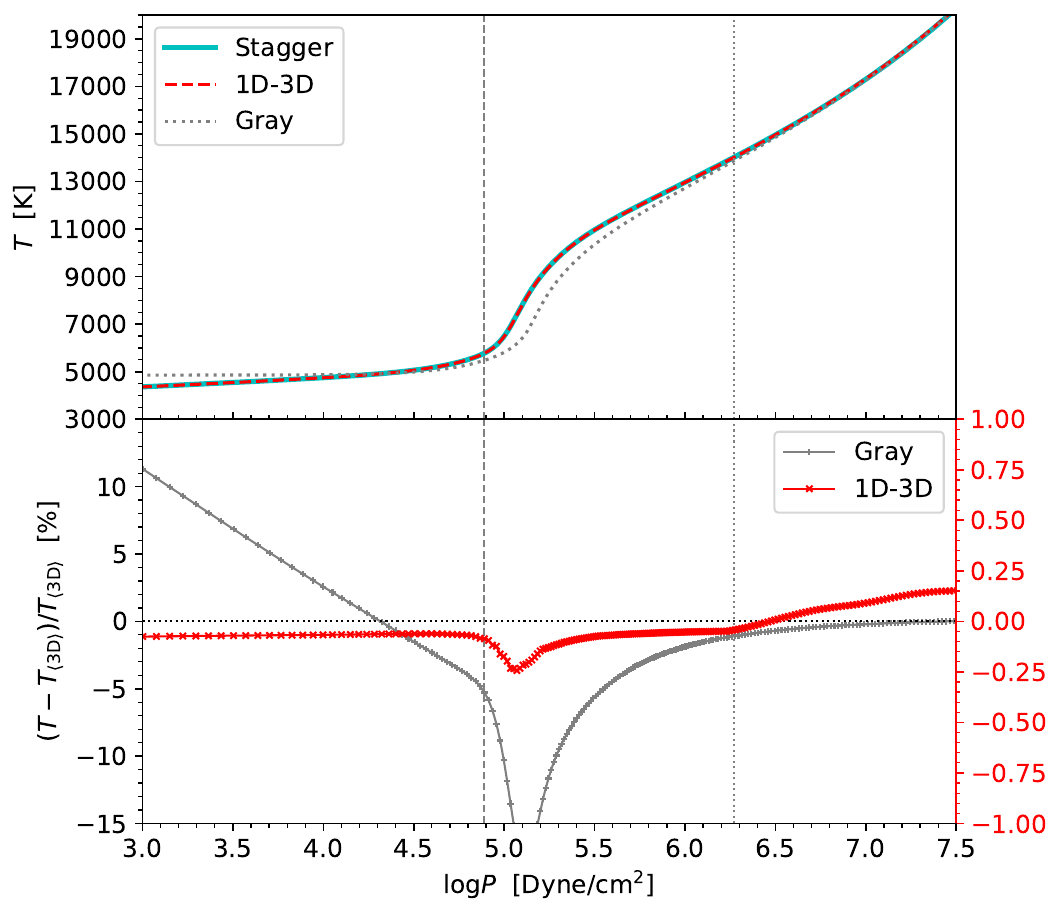}
\caption{\textit{Upper panel:} Temperature profile relative to pressure near the photosphere from the mean \stagger{}-grid solar model (cyan solid line), as well as from \garstec{} solar models calculated with 1D-3D coupling method (red dashed line) and the gray atmosphere (gray dotted line). The vertical dashed line indicates the surface location of the solar model from the 1D-3D coupling method, where temperature equals the solar effective temperature. The matching point (cf.~Fig.~\ref{fig:1D3D-overview}) is marked by the vertical dotted line. 
\textit{Lower panel:} Relative difference (in percentage) between \garstec{} and the mean \stagger{} solar model. Left vertical axis corresponds to temperature differences when the gray atmosphere is employed in the calculation of stellar structure, while a different vertical axis in red is used for the 1D-3D case to make the difference visible.
\label{fig:T-comp}}
\end{figure}

  A basic requirement of all stellar evolution codes is that given appropriate input parameters, evolution calculation must be able to reproduce the properties of the Sun at solar age, 4.57 Gyr. These strict constraints allow us to calibrate free parameters in the evolution code such as $\alpha_{\rm MLT}$, which is referred to as solar calibration. In \garstec{}, solar calibration is carried out as an iterative process. In each evolution iteration, a $1 M_{\odot}$ model is evolved from pre-main-sequence to the solar age with a new combination of mixing-length parameter, initial helium and metal mass fraction. This process is repeated until the modeled luminosity, total radius and mass fraction ratio $Z/X$ agree with the corresponding solar values at solar age within a given error range.
  
  Results of the solar calibration based on the 1D-3D coupling approach are summarized in Table \ref{tb:solar-cali}. The adopted input physics and settings of our evolution calculations are explained in Sect.~\ref{sec:modeling}. We note that considering element diffusion in the construction of solar models improves the agreement between the model-predicted interior structure and that determined from helioseismic data \citep{1993ApJ...403L..75C}; therefore diffusion is taken into account in our solar calibrations and all subsequent evolution calculations, as outlined in Sect.~\ref{sec:1D-model}. The change of surface composition over time is treated more consistently between 1D and \mean{\rm 3D} model due to the newly implemented interpolation in [M/H]. Convective overshoot is not included in the solar calibration, and mass loss is not considered in this work. 
  Meanwhile, as control groups, we carry out solar calibrations for standard evolution calculations adopting the Eddington gray atmosphere, the \citet[KS]{1966ApJ...145..174K} analytical $T(\tau)$ relation fitted to the temperature profile of the solar atmosphere, and the $T(\tau)$ relation fitted to the \citet[their model C, abbreviated to VAL-C]{1981ApJS...45..635V} solar atmosphere \citep{2018A&A...612A..68S} as outer boundary conditions. 
  Additionally, \garstec{} is able to carry out evolution calculations using $T(\tau)$ relations and $\alpha_{\rm MLT}$ calibrated from 3D simulations, as implemented in \citet{2018MNRAS.478.5650M}. We extract the radiative $T(\tau)$ relations from all solar-metallicity \stagger{} grid models according to the formulation of \citet{2014MNRAS.442..805T}, and adopt mixing-length parameters calibrated by \citet{2015A&A...573A..89M}. In this method, the solar calibration sets the scaling factor for the mixing-length parameter, whereas the relative values of $\alpha_{\rm MLT}$ remain unchanged compared to \citet{2015A&A...573A..89M}. The utilization of 3D-calibrated $T(\tau)$ relations and $\alpha_{\rm MLT}$ improves the near-surface temperature stratification throughout the evolution, but we emphasize that it is fundamentally different from the 1D-3D coupling method focused on in this work (see Sect.~\ref{sec:discussion}).
  Input physics of the standard evolution calculations as well as the 3D-calibrated $T(\tau)$ method are identical to the 1D-3D coupling case except for the employed atmospheric boundary condition.
  
  Evolutionary tracks of $1 M_{\odot}$ stars from the zero-age-main-sequence until $\log (g / {\rm [cm/s^2]}) = 2$ with solar-calibrated input parameters are shown in the Kiel diagram (Fig.~\ref{fig:solar-evol}) for these five outer boundary conditions. An additional track adopted from \citet{2018ApJ...869..135S} is shown in the \textit{right panel} of Fig.~\ref{fig:solar-evol}, which was obtained from the entropy calibration method developed in the same study. \citet{2018ApJ...869..135S} adjusted $\alpha_{\rm MLT}$ at every time step of stellar evolution so that the adiabatic entropy of the structure model matches that given by the \stagger{} grid models (see also Sect.~\ref{sec:discussion}).
    The effect of the atmospheric conditions is most pronounced on the RGB, where surface temperatures predicted by the KS atmosphere are about 100 K hotter than the case of the gray atmosphere (see also \citealt{2018ApJ...860..131C}). The evolutionary track given by the 1D-3D coupling method agrees reasonably well with that computed using the gray atmosphere and the 3D-calibrated $T(\tau)$ method with less than 25 K discrepancy in $T_{\rm eff}$ throughout the evolution, albeit a few slight bends are seen in the RGB of the 1D-3D track which will be discussed in Sect.~\ref{sec:star-evol}.
    However, the track from the entropy calibration method differs from the 1D-3D coupling result at the main-sequence-turn-off and by up to about 80 K along the RGB, which deserves further investigation as both methods rely on the same 3D simulations. Implementing both methods to the same evolution code is necessary for a convincing identification of the source of the this difference.
  
  A notable result from Table \ref{tb:solar-cali} is that the mixing-length parameter calibrated with the 1D-3D coupling method is significantly larger than values from standard evolution calculations. According to MLT, $\alpha_{\rm MLT}$ directly impacts the temperature gradient in the convective region. In standard evolution calculations where the model reaches the photosphere, the calibrated $\alpha_{\rm MLT}$ governs the difference between $T_{\rm eff}$ and temperature at the base of the surface convection zone.
  However, in our 1D-3D coupling approach, $\alpha_{\rm MLT}$ is calibrated to reproduce temperature at the matching point of the mean 3D simulation in which convection is modeled in a fundamentally different approach than MLT. Given that (1) the matching temperature $T_{\rm m,\mean{3D}}$ differs from temperature predicted by standard evolution calculations at the corresponding location (see Fig.~\ref{fig:T-comp}), and (2) the superadiabatic temperature gradient at and below the surface and the matching point behaves differently (Fig.~\ref{fig:sad}), mixing-length parameters calibrated from the two approaches are not comparable.\footnote{\citet{2011ApJ...731...78T} defined the mass mixing length in the context of hydrodynamics as the turnover distance of convective upflow (or equivalently, downflow). They quantified the ratio between the mass mixing length, being the ``scale height'' of the mass flux of upflow, and the pressure scale height across their simulation domain and found that the mass mixing-length parameter in the near-adiabatic convective region is greater than its surface value. A higher $\alpha_{\rm MLT}$ calibrated from the 1D-3D coupling method appears qualitatively in line with results from 3D hydrodynamical simulations, although the two mixing-length parameters arise from distinct theories of convection and hence are not directly comparable.}
  Similar discussions can be found in Sect.~3 of paper II.
  
  One of the major advantages of the 1D-3D coupling method is that it makes evolution calculations insensitive to $\alpha_{\rm MLT}$. As shown in Fig.~\ref{fig:alphaMLT}, a 20\% relative change on $\alpha_{\rm MLT}$ shifts $T_{\rm eff}$ by less than 30 K at all evolutionary stages. For standard evolutionary calculations, the same relative change alters $T_{\rm eff}$ by more than 150 K on the main-sequence and more than 200 K along the RGB. As discussed above and seen in Fig.~\ref{fig:sad}, convection is efficient underneath the matching point and convective fluxes are transported via a small $\nabla_T - \nabla_{\rm ad}$. Adjusting $\alpha_{\rm MLT}$ therefore has a marginal effect on the temperature at the outer boundary of stellar interior models. In the usual scenario where the stellar structure model extends to the photosphere, varying $\alpha_{\rm MLT}$ by the same magnitude leaves a larger imprint on the effective temperature and near-surface layers, due to the strong superadiabaticity there (Fig.~\ref{fig:sad}).
  Although the mixing-length parameter is still necessary for modeling convection in stellar structure calculations, it is regarded as constant and fixed to the solar-calibrated value in our approach. We note that observational uncertainties of most stars surpass variations of stellar parameters caused by changing $\alpha_{\rm MLT}$, therefore the 1D-3D coupling method practically makes stellar evolution independent of the mixing-length parameter.

  The near-surface structure of our \garstec{} solar models is compared with the mean 3D \stagger{}-grid solar model in Fig.~\ref{fig:T-comp} (see also paper I Figs.~2-4 and paper II Fig.~5). It is clear that temperature of the 1D-3D coupled model is continuous at the matching point, and the coupled model nearly reproduces the temperature profile of the mean 3D model with less than 0.15\% disagreement in the optically thick regime, which validates the technical aspect of the 1D-3D coupling scheme. The remaining discrepancy in $T$ is due to slight offsets in global parameters between the 1D and 3D model.
  More realistic near-surface structures lead to more accurate theoretical oscillation frequencies. Indeed, as detailed in papers I and II, a major advantage of the 1D-3D coupling method is that the resulting stellar models predict oscillation frequencies in better agreement with measured values, making this novel method suitable for modeling stars with asteroseismic data (cf.~Sect.~\ref{sec:validation}, paper II, and \citealt{2021MNRAS.500.4277J}).

\subsection{Stellar evolution at different metallicities} \label{sec:star-evol}

\begin{figure}[t]
\centering
\subfigure{\includegraphics[width=\columnwidth]{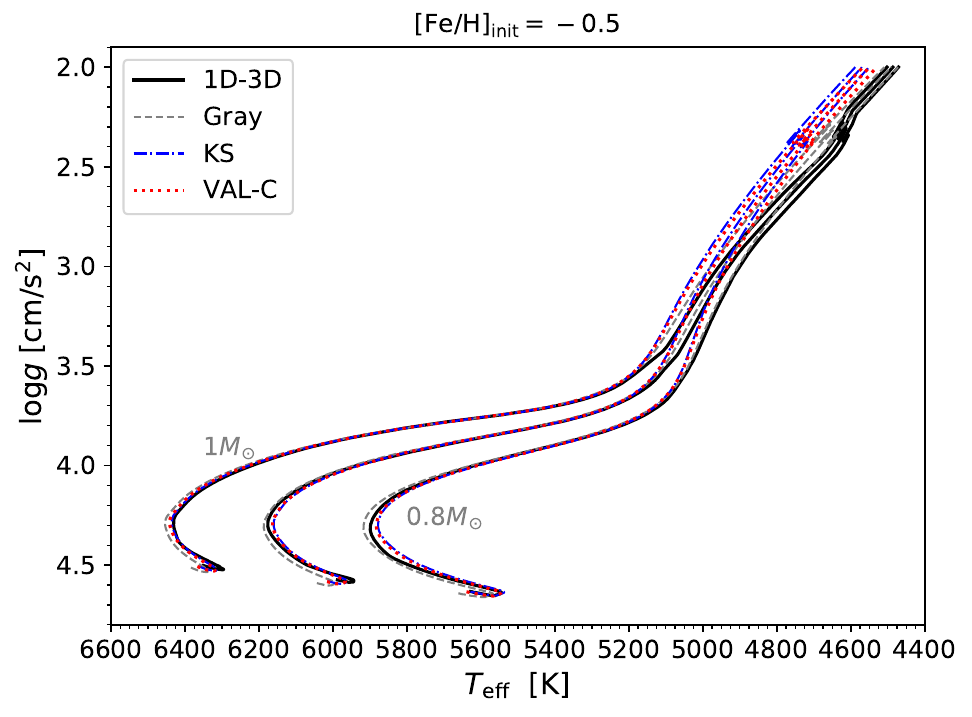}}
\vspace{-2em}%

\subfigure{\includegraphics[width=\columnwidth]{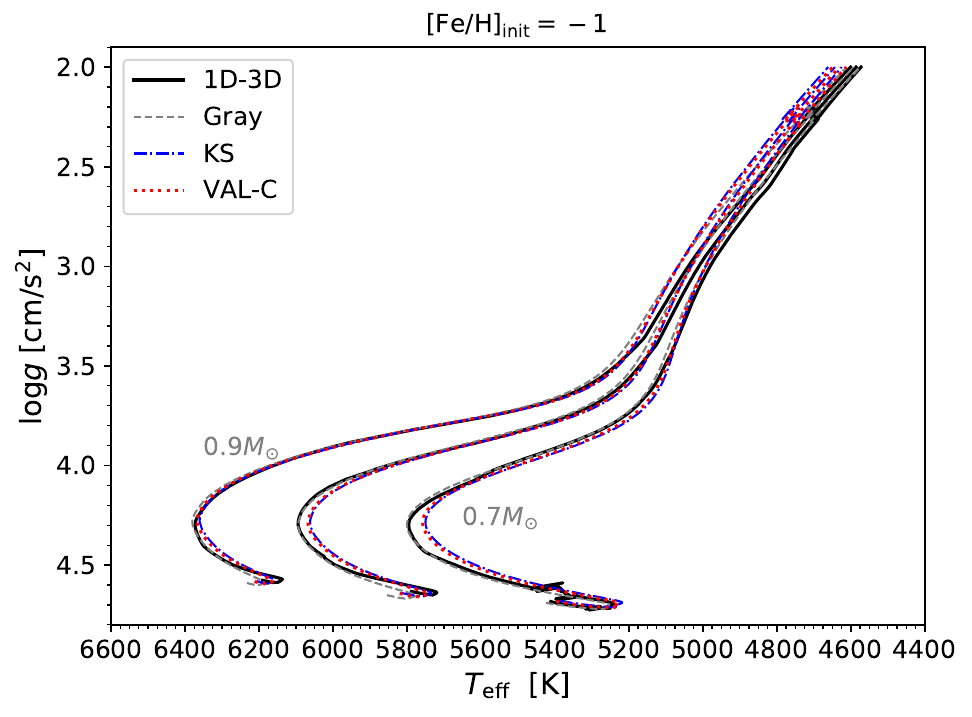}}
\vspace{-2em}%

\subfigure{\includegraphics[width=\columnwidth]{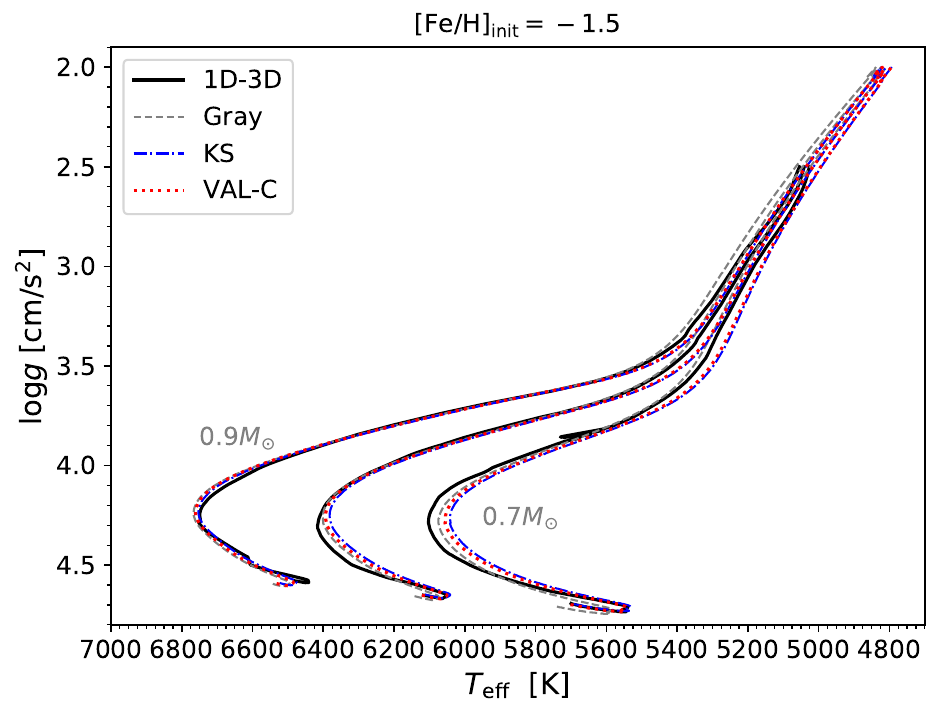}}
\vspace{-2em}%
\caption{Evolutionary tracks at different stellar masses and metallicities computed based on the 1D-3D coupling method, as well as three atmospheric boundary conditions frequently used in stellar evolution calculations. For models with initial metallicity $\rm [Fe/H]_{init} = -1$ and $-1.5$, a 0.4 dex enhancement of $\alpha$-element abundance is applied in \garstec{} calculations, consistent with the $\alpha$-enhanced \stagger{} models when $\rm [Fe/H] \leq -1$. Owing to missing 3D models at $T_{\rm eff} = 5000$ K, $\log g = 2$ (cf.~Fig.~\ref{fig:grid}), $\rm [Fe/H]_{init} = -1.5$ tracks are stopped at $\log g = 2.5$ to avoid extrapolations.}
\label{fig:star-evol}
\end{figure}

  The wide metallicity range covered by the \stagger{}-grid and the interpolation of mean 3D models across $\rm [M/H]$ implemented in this work allow us to extend the 1D-3D coupling method to the evolution of stars with different metallicities. Particularly interesting targets are metal-poor stars which are important in Galactic archaeology, but their fundamental properties predicted by standard evolution calculations sometimes contradict values measured from observations (e.g., \citealt{2015A&A...575A..26C,2024A&A...689A.243C,2018ApJ...856...10J}, but see \citealt{2024ApJ...975...19H} for evidence supporting the reliability of stellar models at low metallicity).
  
  Evolutionary tracks for various masses and initial chemical compositions are shown in Fig.~\ref{fig:star-evol}. All calculations employ their respective solar-calibrated $\alpha_{\rm MLT}$, and other input physics are identical to the solar case. For a given initial [Fe/H], the helium mass fraction of the starting model is determined via $Y_{\rm init} = Y_0 + (\Delta Y / \Delta Z) Z_{\rm init}$, where $Y_0$ is set to 0.248. The helium enrichment coefficient $\Delta Y / \Delta Z = 0.97$ is fixed by solving this equation with $Y_{\rm init}$ and $Z_{\rm init}$ obtained from solar calibration (cf.~Table \ref{tb:solar-cali}). An investigation of the helium enrichment law is beyond the scope of this paper. 

  It is worth noting that for warmer stars whose $T_{\rm eff} \gtrsim 6300$ K, including the standard treatment of element diffusion results in significant or complete depletion of helium and metal by gravitational settling (see Fig.~1 of \citealt{2019MNRAS.489.1850V}). Heavier elements are easily sedimented in A and F-type stars due to their thin surface convection zone and large pressure gradient near the surface. Such surface abundance patterns predicted by standard diffusion calculations are not in line with helium and metal abundances measured in F-type stars \citep{1999A&A...351..247V,2019MNRAS.483.4678V}. 
  Since the metallicity interpolation in our 1D-3D coupling method is based on $Z/X$, the strong depletion of metals when $T_{\rm eff} \gtrsim 6300$ K is also a practical problem that results in the interpolation scheme picking up unrealistically metal-poor \mean{\rm 3D} models or even going out of the metallicity range.
  
  To this end, we implemented the so-called turbulent diffusion in our calculations to counter the effect of gravitational settling. Turbulent diffusion was put forward as an extra mixing term to reproduce the surface abundance of lithium in warm population II stars \citep{1991ApJ...371..584P,2005ApJ...619..538R} and the abundance pattern of turn-off stars in metal-poor globular clusters \citep{2012ApJ...753...48N}. It aims to represent the effect of turbulence that might be generated by mild convective overshooting in the radiative regions below the surface convection zone, as stated in \citet{1991ApJ...371..584P}. Here the turbulent diffusion coefficient is evaluated following Eq.~1 of \citet{2017ApJ...840...99D}:
\begin{equation}
D_T = D_0 \left( \frac{\rho_{\rm CZ}}{\rho} \right)^3 \left( \frac{M_{\rm CZ}}{M} \right)^{-3/2},
\end{equation}
where $\rho_{\rm CZ}$ and $M_{\rm CZ}$ are density at the base of the surface convection zone and the total mass of the surface convection zone, respectively. The free parameter $D_0$ is set to $1 \rm cm^2/s$ \citep{2017ApJ...840...99D}. We apply turbulent diffusion at and under the base of the surface convection zone. If convective overshoot is also included in stellar structure modeling, $D_T$ is added on top of the diffusion coefficient from overshooting. We emphasize that although turbulent diffusion is conceived based on reasonable assumptions and motivations, it does not have a solid physical basis. Radiative acceleration and rotationally induced mixing are physically well-justified processes that counteract gravitational settling; however it is challenging to incorporate them into stellar evolution calculations (see \citealt{1998ApJ...492..833R} and \citealt{2020A&A...633A..23D} for efforts in this topic).

  Fig.~\ref{fig:star-evol} shows how surface boundary conditions affect the evolution of metal-poor stars. At all metallicities,\footnote{We have checked that different treatments of the surface layer hardly affect element diffusion and the evolution of surface abundance. This is also evidenced from the solar calibration that initial and modeled present-day $X$, $Y$, $Z$ values are insensitive to the outer boundary condition used (Table \ref{tb:solar-cali}).} evolutionary tracks computed with the 1D-3D coupling approach are similar to that given by the gray atmosphere. Less than 25 K differences in $T_{\rm eff}$ between the two types of evolution calculations are found at the main-sequence turn-off, while on the RGB the discrepancy depends on metallicity and the evolutionary stage but does not exceed 40 K in all cases investigated. Overall, the 1D-3D coupling method gives cooler tracks along the RGB than the gray atmosphere scenario.
  On the other hand, metal-poor tracks computed from the KS and VAL-C atmosphere resemble each other closely, but they are clearly distinguishable from the 1D-3D coupling result. 
  It is worth noting that we find no physical explanation for the similarity in evolutionary tracks between the 1D-3D coupling approach and the gray atmosphere scenario. The latter adopted unrealistic assumptions in stellar atmospheres and predicted significantly different $T(\tau)$ relations than 3D model atmospheres (see Fig.~2 of \citealt{2024ApJ...962..118Z}). Moreover, the treatment of the outer boundary and surface convection is fundamentally different between the standard evolution calculation and the 1D-3D coupling method. Therefore, we speculate the observed reasonable agreement is a pure coincidence -- the results seen in Fig.~\ref{fig:star-evol} do not imply that the gray atmosphere is more realistic than KS or VAL-C, nor can they be interpreted as evidence to support the use of the gray atmosphere boundary conditions in standard evolution calculations.
  
  Another notable aspect of the 1D-3D coupling method is that the computed tracks demonstrate discontinuities and bends around certain stellar parameters. This can be seen from zigzags around $T_{\rm eff} = 5400$ K, $\log (g / {\rm [cm/s^2]}) = 4.65$ in the middle panel and around $T_{\rm eff} = 5700$ K, $\log (g / {\rm [cm/s^2]}) = 3.85$ in the bottom panel of Fig.~\ref{fig:star-evol}. Bends along the RGB, which were referred to as ``kinks'' and discussed in detail in paper II, are visible in Figs.~\ref{fig:grid}, \ref{fig:solar-evol} and \ref{fig:star-evol}.
  Discontinuities in the tracks are certainly unphysical and stem from numerical issues. They are diagnosed to be mainly associated with problems in partial derivatives at the outermost shell of the \garstec{} model. Mean 3D models not only supply pressure and luminosity at the outer boundary but also their partial derivatives with respect to temperature and radius, which are evaluated numerically as well. As derivatives from 2D linear interpolation are not guaranteed to be continuous, large jumps in pressure and/or luminosity derivatives may occur when global parameters of the star vary across the boundary of two triangulations defined in the interpolation scheme. Large variations in partial derivatives will result in the Henyey solver converging to a new solution that is far away from the previous time step. 
  Bends along RGB tracks appear from similar underlying reasons. Although the 2D cubic interpolation of \citet{Renka1984} employed during the red-giant evolution is $C^1$ continuous, i.e., continuous first derivatives, we observe rapid change of derivatives at the parameter space where the bend (or change of direction) takes place.
  Apart from making the 3D grid denser as advocated in papers I and II, which will likely reduce sudden variations during interpolation, a suitable interpolation method and carefully evaluated partial derivatives are also necessary to make evolutionary tracks smoother. We will focus on these two aspects for further improvements on the 1D-3D coupling method in the future.

\section{Validation against binary systems} \label{sec:validation}

  Detached binary systems are ideal candidates to test stellar evolution calculations. It is commonly believed that two components are born in the same epoch from a common molecular cloud (see \citealt{2024A&A...682L..23S}, however, for evidence pointing to a different scenario) thus should have identical age and initial chemical composition. This characteristic, combined with accurate stellar mass and radius determined from orbits, chemical abundance from spectroscopy, and luminosity or effective temperature from photometry, put strong constraints on stellar evolution modeling. Moreover, oscillating stars in binary systems, which stand on the intersection of two effective approaches for accurate determination of stellar parameters, are valuable targets for calibrating adjustable parameters that govern convection (e.g., \citealt{2018MNRAS.475..981L,2024ApJ...974...77L}) and internal mixing \citep{2016A&A...592A.116S,2019MNRAS.482.1231J} in 1D stellar modeling, and for testing the asteroseismic scaling relation \citep{2016AN....337..793B,2016ApJ...832..121G}. Here we validate the 1D-3D coupling method against both components in the eclipsing binary AI Phe and two oscillating giants in binary systems observed by \textit{Kepler}, KIC 9970396 and KIC 10001167. Given that stellar evolution using the 1D-3D coupling method is insensitive to the mixing-length parameter (Fig.~\ref{fig:alphaMLT}), we fix $\alpha_{\rm MLT}$ to the solar-calibrated value throughout. Having invariant $\alpha_{\rm MLT}$ and tight constraint on stellar mass, the only major tunable input parameter is the initial chemical composition $X$, $Y$ and $Z$. With very few ``degrees of freedom'' in the evolution calculation, we demonstrate below that our models reproduce various observables reasonably well.

\subsection{AI Phe}

\begin{table*}
\centering
\caption{Fundamental parameters of the binary system AI Phe from observational data and evolution modeling using the 1D-3D coupling method.
\label{tb:AIPhe}}
\begin{threeparttable}
{\begin{tabular*}{\textwidth}{@{\extracolsep{\fill}}ccccccc}
\toprule[2pt]

 \multicolumn{2}{c}{Stellar parameter} & $M/M_{\odot}$ & $R/R_{\odot}$ & $T_{\rm eff}$ [K] & [Fe/H] & Age ($10^9$yr)  \\

\midrule[1pt]

\multirow{2}{*}{Primary}  
	& Observation 	& $1.2438 \pm 0.0008$ (a)	& $2.9303 \pm 0.0023$ (a)	& $5094 \pm 50$ (b)	& $-0.13 \pm 0.1$ (c)	& --- \\
 
 	& Modeling 		& $1.244$ 					& $2.934$ 					& $5047$  			& $-0.10$    			& 4.73 \\
 
\midrule[1pt]

\multirow{2}{*}{Secondary}  
	& Observation 	& $1.1938 \pm 0.0008$ (a)	& $1.8036 \pm 0.0022$ (a)	& $6199 \pm 50$ (b)	& $-0.17 \pm 0.1$ (c)	& --- \\
 
 	& Modeling 		& $1.194$ 					& $1.803$ 					& $6177$   			& $-0.22$     			& 4.69 \\
 
\bottomrule[2pt]
\end{tabular*}}

	\begin{tablenotes}
    	\item Reference: (a) \citet{2020MNRAS.498..332M}; 
    	(b) \citet{2020MNRAS.497.2899M}, uncertainties quoted from \citet{2023A&A...678A.203V};
    	(c) \citet{1988A&A...196..128A}                
    \end{tablenotes}
    
\end{threeparttable}
\end{table*}

\begin{figure}
\centering
\includegraphics[width=\columnwidth]{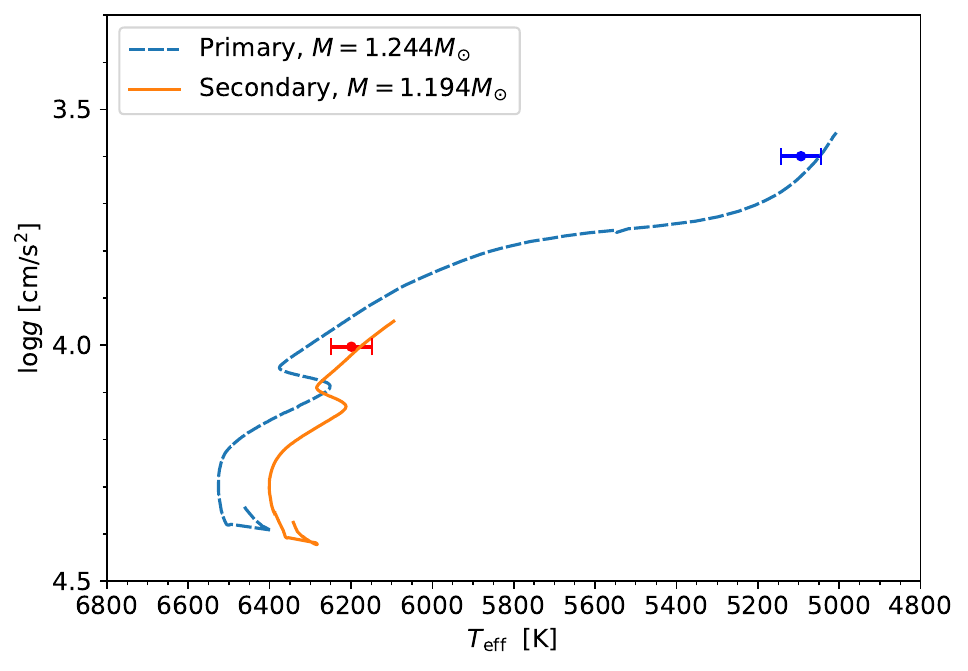}
\caption{Evolutionary tracks computed using the 1D-3D coupling method, which are good representations of the primary and secondary component of AI Phe, are compared with the corresponding observational data (blue and red dots with error bars). Stellar parameters of the best-matching model are listed in Table \ref{tb:AIPhe}.
\label{fig:AIPhe}} 
\end{figure}

  AI Phe (HD 6980) is an eclipsing binary system composed of a subgiant and a more evolved star at the base of RGB. As a bright target, it has been studied in detail observationally by, among others, \citet{1978IBVS.1419....1R} and \citet{2020MNRAS.498..332M}, making it an important benchmark for testing stellar evolution models \citep{2013ApJ...776...87S,2017A&A...608A..62H,2023A&A...678A.203V}. We adopt stellar masses and radii measured by \citet{2020MNRAS.498..332M}, who carefully determined geometric and orbital parameters of AI Phe from TESS light curves as well as spectroscopic orbits. Fundamental stellar parameters of each component and their sources are tabulated in Table \ref{tb:AIPhe}.
  
  In our modeling, we require the two components to have identical initial chemical composition, and the difference in their age must be less than 1\%. For each star, the model-predicted global properties, that is, radius, $T_{\rm eff}$ and surface [Fe/H], should match the measured values within their uncertainties. 
  A grid of evolutionary tracks is computed for each component with fixed stellar mass due to the strong observational constraint. We vary initial [Fe/H] from -0.24 to -0.04 dex, initial helium mass fraction from 0.255 to 0.275, and the overshoot parameter in the stellar structure calculation, $f_{\rm ov}$ (cf.~Sect.~\ref{sec:1D-model}), from 0 (no overshoot) to 0.03. Overshooting above the surface convection zone is not modeled by \garstec{}, so $f_{\rm ov}$ has no effect in that region.
  
  Given these requirements and relatively few adjustable parameters, the 1D-3D coupling method is able to produce acceptable models for AI Phe. A representative track for both components is shown in the Kiel diagram in Fig.~\ref{fig:AIPhe}, and the corresponding modeled stellar parameters are listed in Table \ref{tb:AIPhe}. 
For both components, the modeled $T_{\rm eff}$ is in reasonable agreement with measured values. Considering the statistical uncertainties given by \citet{2020MNRAS.497.2899M} and \citet{2023A&A...678A.203V} and in view of additional systematic errors in the determination of effective temperature, the less than 50 K discrepancy between evolutionary models and observations is acceptable. Another useful perspective is to examine the $T_{\rm eff}$ ratio of the two components, a quantity directly determined from the observed light curves \citep{2017A&A...608A..62H}. The modeled temperature ratio between the primary and secondary component is 0.817, consistent with the measured value $0.821 \pm 0.004$ \citep[their Table 5]{2020MNRAS.497.2899M} within uncertainties.
  The predicted age of the binary system is about 4.7 Gyr, which agrees well with detailed modeling results of \citet{2017A&A...608A..62H} and \citet{2023A&A...678A.203V}.
  
  Additionally, as the metallicity of the primary and secondary component of AI Phe are determined with the same method and observational data \citep{1988A&A...196..128A}, the measured metallicity difference, $\Delta\rm [Fe/H]_{obs} = 0.04$, should be realistic. This is analogous to the differential abundance analysis for solar twins that leads to extremely precise element abundance for stars similar to the Sun (e.g.~\citealt{2014ApJ...791...14M}). 
  However, our modeling yields $\Delta\rm [Fe/H]_{mod} = 0.12$. The significant discrepancy in $\Delta\rm [Fe/H]$ indicates room for improvement in the treatment of element diffusion, including the turbulent diffusion, in our modeling.

\subsection{KIC 9970396} \label{sec:KIC9970}

\begin{table*}
\centering
\caption{Fundamental parameters of KIC 9970396 from observations, the best-fitting model, and median values together with the 16\% and 84\% limit of the inferred posterior distribution (the 68\% Bayesian credibility interval) from \texttt{BASTA}. The frequency of maximum oscillation power, $\nu_{\max}$, of the models is estimated from the asteroseismic scaling relation \citep{1991ApJ...368..599B,1995A&A...293...87K}.
\label{tb:KIC9970}}
\begin{threeparttable}
{\begin{tabular*}{\textwidth}{@{\extracolsep{\fill}}cccccccc}
\toprule[2pt]

  Stellar parameter & $M/M_{\odot}$ & $R/R_{\odot}$ & $L/L_{\odot}$ & $T_{\rm eff}$ [K] & [Fe/H] & $\nu_{\max} \; \rm [\mu Hz]$ & Age ($10^9$yr)  \\

\midrule[1pt]

  Observation 	& $1.178 \pm 0.015 $ (a)	& $8.035 \pm 0.074 $ (a)	& $30.53 \pm 1.85$ (b) & $4825 \pm 85$ (c)	& $-0.27 \pm 0.08$ (c)  & $63.8 \pm 0.5$ (d)        & --- \\
 
  Best model      & $1.229$ 				    & $8.052$                   & $30.83$              & $4793$   		    & $-0.36$ 				& $64.5$                    & $4.05$\\
  
  Inferred values & $1.214_{-0.030}^{+0.020}$  & $8.023_{-0.068}^{+0.043}$ & $31.89_{-1.23}^{+0.97}$   & $4852_{-57}^{+39}$    & $-0.46_{-0.05}^{+0.07}$   & $63.6_{-0.4}^{+0.6}$      & $4.04_{-0.22}^{+0.30}$\\
 
\bottomrule[2pt]
\end{tabular*}}

	\begin{tablenotes}
    	\item Note: (a) \citet{2018MNRAS.476.3729B}; 
    	(b) Determined based on $K_s$-band magnitude \citep{cutri2003}, Gaia DR3 distance \citep{bailer-jones2021}, extinction correction \citep{green2019}, and bolometric correction \citep{choi2016}, using the procedure implemented in \texttt{ISOCLASSY} \citep{huber2017,berger2020}. We have accounted for the parallax zero-point offset in the Gaia data \citep{2021A&A...654A..20G} and have inflated uncertainties following the methods proposed by \citet{2021MNRAS.506.2269E};
    	(c) The parameter and its error are averaged from \citet[based on spectra from APOGEE]{2016ApJ...832..121G} and \citet{2018MNRAS.476.3729B};
        (d) \citet{2018MNRAS.475..981L}
    \end{tablenotes}
    
\end{threeparttable}
\end{table*}

\begin{figure*}
\centering
\includegraphics[width=0.7\textwidth]{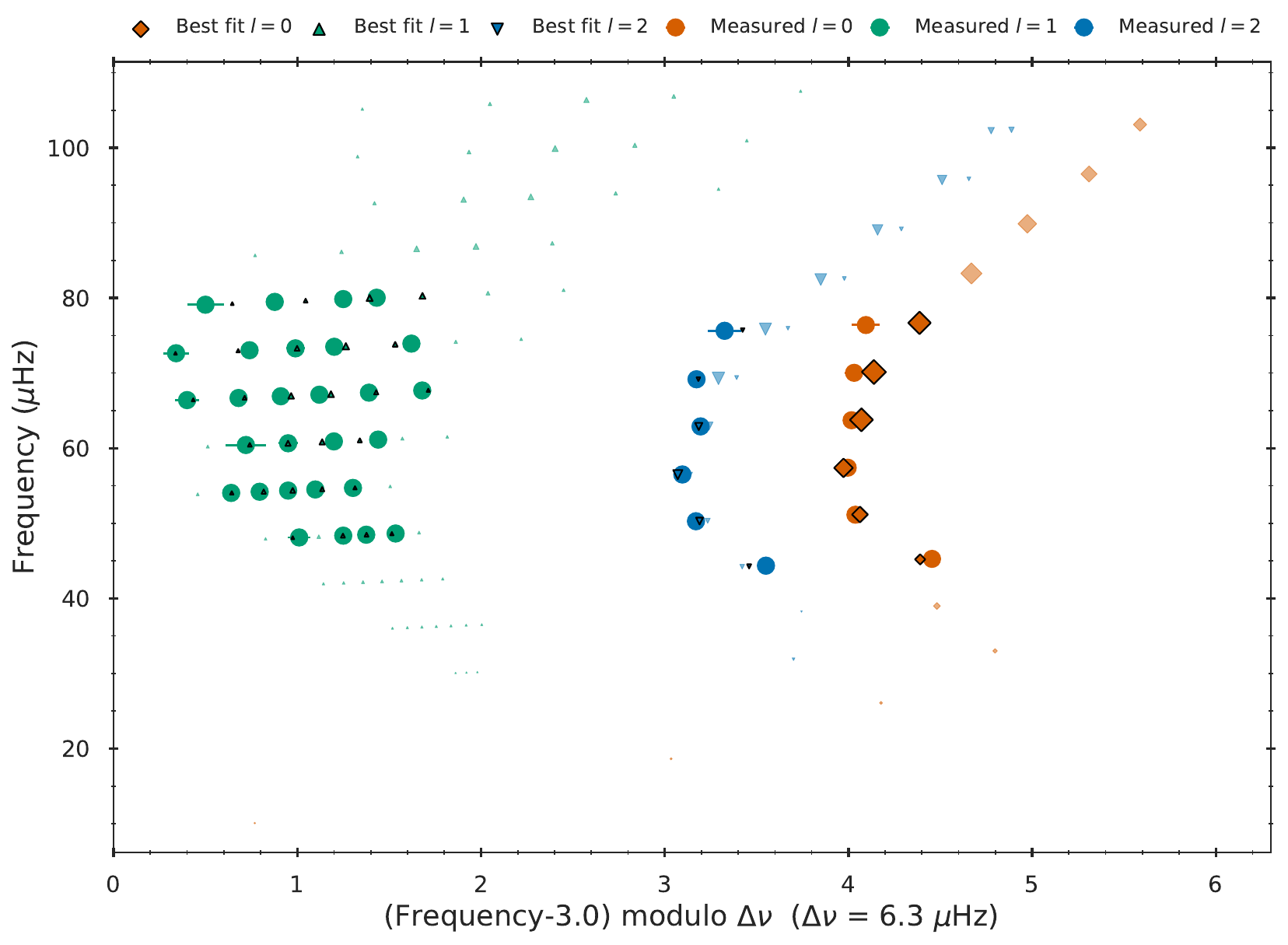}
\caption{{\'E}chelle diagram of the best-fitting model of KIC 9970396 (Table \ref{tb:KIC9970}). Here $\Delta\nu$ is the large frequency separation. The orange, green and blue filled circles indicate frequencies of $l = 0,1,2$ modes measured by \citet{2018MNRAS.475..981L}. Theoretical frequencies calculated from \texttt{ADIPLS} using the truncated scanning method are orange rhombuses, green triangles, and blue inverted triangles whose sizes are inversely proportional to their mode inertias. No frequency correction is applied to the theoretical frequencies. Transparent symbols correspond to theoretical frequencies that are not compared with observations.
\label{fig:KIC9970}} 
\end{figure*}

  KIC 9970396 is a detached eclipsing binary system detected by \textit{Kepler} \citep{2010Sci...327..977B,2011AJ....141...83P}. It consists of an oscillating red giant and a G or late F companion. Global parameters of the binary and key asteroseismic parameters were first provided in \citet{2013ApJ...767...82G,2016ApJ...832..121G} and subsequently investigated in detail by \citet{2018MNRAS.476.3729B}. Furthermore, \citet{2018MNRAS.475..981L} extracted 40 oscillation frequencies of low-degree modes for KIC 9970396 from \textit{Kepler} photometric data. With well-determined stellar mass and radius from the binary orbit and a wealth of mode frequencies that provide strong constraints on the interior structure, KIC 9970396 is an ideal test bed for stellar structure modeling. The red giant was used as an important target in asteroseismic modeling by \citet{2018MNRAS.475..981L} and \citet{2022ApJ...931...64Z} for calibrating the mixing-length parameter and stellar age determination, respectively.
  
  In this work, we aim to construct evolutionary models for KIC 9970396 that reasonably reproduce the observed stellar parameters and, more importantly, oscillation frequencies. We consider not only measured radial modes but also the $l = 1$ and $l = 2$ mixed modes where $l$ is the mode degree. Mixed modes, as explained in \citet{2017A&ARv..25....1H}, are found in subgiant and red-giant stars in which propagation cavities of pressure modes and gravity modes (g-mode) are coupled (or not well separated). Mixed modes have substantial amplitudes in both the outer region and near the stellar core, demonstrating distinctive characteristics different from p- or g-modes. Incorporating mixed modes in our analysis provides unique constraints on the core properties and evolutionary stage of our model.
  
  Similar to the case of AI Phe, we computed a grid of evolutionary tracks with different stellar masses, initial metallicities and amount of convective overshoot in stellar interior models. Stellar structural models are selected once the modeled stellar radius approaches the measured value (cf.~Table \ref{tb:KIC9970}). We subsequently compute theoretical frequencies for all selected models using the Aarhus adiabatic oscillation package (\texttt{ADIPLS}, \citealt{2008Ap&SS.316..113C}). It is worth noting that for red giants, numerical calculations predict numerous theoretical mixed modes that could potentially be excited. This makes the calculation of the frequencies for the stellar models computationally demanding especially in the context of grid-based modeling. Nevertheless, only mixed modes with low inertia, that is, oscillations that involve relatively small fraction of the stellar mass (often referred to as p-dominated mixed modes), are observable from power spectra (e.g., Fig.~5 of \citealt{2018MNRAS.475..981L}) since they have a higher amplitude at a given excitation and damping rate.
  To facilitate straightforward comparison between theoretical and measured mixed mode frequencies, we employ the \textit{truncated scanning method} recently developed by \citet{2024A&A...690A.394L}. As evident from its name, the method utilizes a truncated stellar model with the innermost region removed to estimate the acoustic resonances (the most p-mode-like mixed modes whose frequencies approximately satisfy the acoustic asymptotic relation) within each mode order. They are then used as centerpoints of the observable frequency ranges, for which the complete structure model is subsequently scanned to obtain the observable p-dominated mixed modes.
  The truncated scanning method outputs much fewer mixed mode solutions, hence significantly reducing the computation cost.
  
  All selected stellar structure models and their corresponding oscillation frequencies computed with the truncated scanning method are subsequently fed into the BAyesian STellar Algorithm (\texttt{BASTA}; \citealt{2015MNRAS.452.2127S,2022MNRAS.509.4344A}) to find the model that best represents KIC 9970396. Based on Bayesian statistics, \texttt{BASTA}\footnote{\url{https://github.com/BASTAcode/BASTA}} is a versatile fitting algorithm that infers stellar properties using a given set of observational data which could stem from astrometry, spectroscopy, asteroseismology or combinations thereof. Meanwhile, it is highly flexible in terms of stellar tracks or isochrones adopted in the fitting. In the case of KIC 9970396, the observed radius, effective temperature, as well as all individual frequencies extracted in \citet[their Table 3]{2018MNRAS.475..981L} are fitted to the models. For matching the observed individual frequencies with corresponding frequencies in the model, we use the default algorithm in \texttt{BASTA} described in \citet{2019MNRAS.489..928S} and \citet{2022MNRAS.509.4344A}\footnote{A detailed description of the frequency-fitting algorithm used in \texttt{BASTA} will be presented in Stokholm et al.~(in preparation).}.
  
  The best-fitting model is defined as the one with maximum likelihood, whose fundamental parameters are listed in Table \ref{tb:KIC9970} and theoretical frequencies are compared against observations through the {\'e}chelle diagram shown in Fig.~\ref{fig:KIC9970}. To avoid seismic measurements dominating the likelihood, $\chi^2$ computed from seismic data is divided by the total number of frequencies (see \citealt{2022MNRAS.509.4344A} Eq.~4).
  We note that no frequency correction for the asteroseismic surface effect is applied in the fitting process, as (1) the 1D-3D coupling approach greatly reduces the discrepancy between the theoretical and measured frequencies (cf.~paper I Fig.~5 and paper II Fig.~10) and (2) the remaining surface terms have a different shape than those from traditional models. The median, 16th, and 84th quantiles of posterior distributions of stellar parameters inferred based on our model grid for KIC 9970396 are also listed in Table \ref{tb:KIC9970}.
  
  The {\'e}chelle diagram depicts the relationship between mode frequency and $\nu$ mod $\Delta\nu$, where $\Delta\nu$ is the large frequency separation defined as the frequency difference between two modes with the same degree and consecutive radial order. The diagram divides the frequency spectrum into sections so that p-modes with the same $l$ line up nearly vertically in the diagram whereas modes with different degrees are well separated, thus offering a clear manner to compare theoretical and observed frequencies. As seen in Fig.~\ref{fig:KIC9970}, theoretical frequencies predicted from our best model agree reasonably well with observation for both radial p-modes and non-radial mixed modes. Systematic offsets are still present at high frequencies, which is expected as the 1D-3D coupling method is known to reduce but not eliminate the asteroseismic surface effect. Although the mixing-length parameter is not adjusted within our approach, we are able to model oscillation frequencies of KIC 9970396 at a similar level of fidelity as the detailed investigation by \citet[see their Fig.~A6; after applying surface corrections they achieve better agreement at high frequencies]{2018MNRAS.475..981L}. The stellar mass inferred from our modeling falls between the dynamical mass and the higher values determined from the (corrected) asteroseismic scaling relation (see \citealt{2018MNRAS.476.3729B} Fig.~7). Given that the global properties of our model are also compatible with measured values, we are confident that the 1D-3D coupling method produces reliable structural models for KIC 9970396 with realistic near-surface stratification, which is an encouraging validation of our method at non-solar metallicity.

\subsection{KIC 10001167}  \label{sec:KIC1000}

\begin{table*}
\centering
\caption{Fundamental parameters of KIC 10001167 from observations, the best-fitting model and median values together with the 16\% and 84\% limit of the inferred posterior distributions.
\label{tb:KIC1000}}
\begin{threeparttable}
{\begin{tabular*}{\textwidth}{@{\extracolsep{\fill}}ccccccccc}
\toprule[2pt]

  Stellar parameter & $M/M_{\odot}$	& $R/R_{\odot}$	& $L/L_{\odot}$	& $T_{\rm eff}$ [K]	& [Fe/H]		& $\rm [\alpha/Fe]$	& $\nu_{\max} \; \rm [\mu Hz]$	& Age ($10^9$yr)  \\

\midrule[1pt]

  Observation 	& $0.934 \pm 0.008$ (a)	& $13.03 \pm 0.21$ (a)	& $71.7 \pm 3.8$ (b)		& $4625 \pm 59$ (a)	& $-0.73 \pm 0.10$ (a)	& $0.37 \pm 0.10$ (a)	& $19.9 \pm 1.0$ (c)	& --- \\
  
  Best model		& $0.998$				& $13.03$ 				& $69.2$ 			 & $4618$    			& $-0.82$				& $0.4$					& $20.4$					& $7.57$\\

  Inferred values & $0.969_{-0.030}^{+0.025}$	& $12.90_{-0.15}^{+0.10}$                & $66.5_{-3.6}^{+3.9}$	& $4603_{-62}^{+54}$	& $-0.75_{-0.10}^{+0.10}$ & --- & $20.2_{-0.2}^{+0.2}$	& $8.60_{-0.79}^{+1.24}$\\
 
\bottomrule[2pt]
\end{tabular*}}

	\begin{tablenotes}
    	\item Note: (a) \citet{2025arXiv250417853T}; 
    	(b) cf.~note (b) in Table \ref{tb:KIC9970};
    (c) \citet{2024ApJ...974...77L}
    \end{tablenotes}
    
\end{threeparttable}
\end{table*}

\begin{figure*}
\centering
\includegraphics[width=0.7\textwidth]{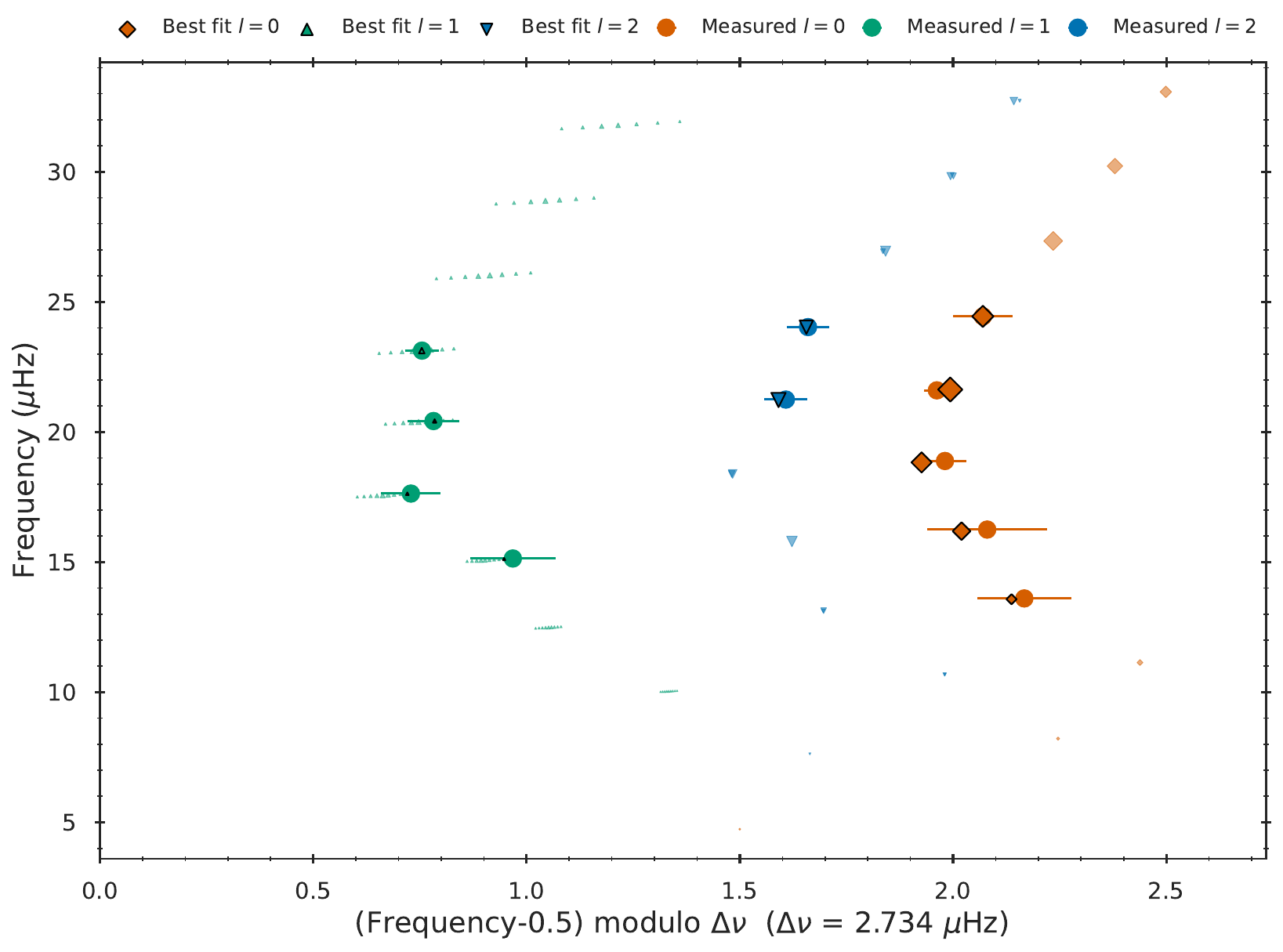}
\caption{{\'E}chelle diagram of the best-fitting model of KIC 10001167 (Table \ref{tb:KIC1000}). Orange, green, and blue filled circles indicate measured frequencies of $l = 0,1,2$ modes. No frequency correction is applied to the theoretical frequencies.
\label{fig:KIC1000}} 
\end{figure*}

  KIC 10001167 is another eclipsing binary system observed by \textit{\hbox{Kepler}}. Similar to KIC 9970396, it is composed of a G or F-type dwarf and a more evolved oscillating red giant \citep{2013ApJ...767...82G}. The system was thoroughly analyzed in a recent study by \citet{2025arXiv250417853T}, who derived precise dynamical mass and radius for the oscillating giant. Its low metallicity, $\rm [Fe/H] = -0.73$ and a 0.37 dex enhancement of $\alpha$-element \citep{2025arXiv250417853T}, makes KIC 10001167 an ideal target for validating the 1D-3D coupling method. The red giant was modeled in detail in \citet{2024ApJ...974...77L} as one of the binary samples to identify the relationship between metallicity and the mixing-length parameter.

  We measure the oscillation frequencies of $l = 0,1,2$ modes following the method described in \citet{litd++24-m67}. In brief, we identify frequency bins with signal-to-noise ratios greater than five as prominent spikes, and then extract a mode where a number of spikes appear within the typical width of an oscillation mode. By visual inspection, we identify the $l = 0, 2$ modes based on their characteristic pair-peak features, and treat the rest as the $l = 1$ modes. We fit a mode with the Lorentz function to measure the frequency and estimate the uncertainty using the method described in \citet{2018MNRAS.475..981L}, who applied the Monte Carlo approach and produced 1000 simulated power spectra by multiplying the power spectrum by a random noise following a $\chi^2$ distribution with 2 degrees of freedom. The standard deviation of the 1000 measured frequencies is used as the uncertainty of a mode.

  A grid of evolutionary tracks with stellar masses ranging from $0.86$ to $1 M_{\odot}$ and initial [Fe/H] from -0.85 to -0.6 is computed to model KIC 10001167. The $\alpha$-enhancement used in our evolutionary grid calculation is 0.3 or 0.4 dex. The frequency calculation and fitting procedure are described in Sect.~\ref{sec:KIC9970}. Theoretical models are fitted to the dynamical radius, effective temperature (cf.~Table \ref{tb:KIC1000}), as well as measured frequencies of $l = 0,1,2$ modes. Classical parameters inferred from our grid-based modeling and parameters of the model with the maximum likelihood are tabulated in Table \ref{tb:KIC1000}, and Fig.~\ref{fig:KIC1000} is the {\'e}chelle diagram of KIC 10001167.
  
  The 1D-3D coupling method is able to produce acceptable models for KIC 10001167, whose radius, effective temperature, and individual frequencies agree reasonably well with corresponding observations. The inferred stellar mass and mass of the best fitting model are greater than that measured from the binary orbit, similar to the case of KIC 9970396. If we instead fit the coupled models to the measured mass, effective temperature and mode frequencies, the derived stellar radius turns out to be $12.73 \pm 0.05 \; R_{\odot}$, being smaller than the dynamical radius. Results from these two scenarios indicate that when both fundamental stellar parameters and oscillation frequencies are used as constraints, the 1D-3D coupling method tends to slightly overestimate surface gravity and mean density ($\bar{\rho}$) along the RGB. This assertion is further supported by the fact that our best model for KIC 10001167 predicts slightly larger $\Delta\nu$ ($\rm 2.767$ versus $2.734 \pm 0.033 \;\rm \mu Hz$ as measured by \citealt{2024ApJ...974...77L}) and $\nu_{\max}$. These two key asterosesmic observables are proportional to $\sqrt{\bar{\rho}}$ and $g / \sqrt{T_{\rm eff}}$, respectively, according to the asteroseismic scaling relations. 
  The underlying reason for this offset is not understood\footnote{We note that \citet{2025arXiv250417853T} modeled KIC 10001167 with traditional evolutionary calculations and also predicted a larger surface gravity and mean density compared with observations (see their Table G.2).}. One possible cause from the modeling aspect could be the remaining small surface effects from coupled models (e.g.~Fig.~\ref{fig:KIC1000} and Fig.~5 of paper I) that were not accounted for in our fitting process, which might lead to systematics in derived stellar parameters. A detailed study of asteroseismic surface effects in red giants using coupled or patched models whose near-surface stratification is supplied from 3D simulations is needed to shed light on this issue.

\section{Discussion: different approaches of improving stellar evolutionary models with 3D surface convection simulations} \label{sec:discussion}

  As outlined in the introduction, considerable effort has been devoted to overcoming the outstanding problem in stellar interior modeling, where phenomenological theories of convection such as the MLT fail in the near-surface convective layers. To our knowledge, there are three well-established methods that utilize 3D model atmosphere grids to improve the treatment of near-surface layers in stellar evolution calculations. Apart from the 1D-3D coupling method developed and tested in papers I, II, and this work, \citet{2014MNRAS.442..805T,2014MNRAS.445.4366T} carefully extracted the $T(\tau)$ relation from the \citet{2013ApJ...769...18T} grid and performed a calibration of $\alpha_{\rm MLT}$ based on the same 3D grid; \citet{2018ApJ...869..135S,2021MNRAS.504.3128S} and \citet{2024A&A...687A.146M} used the adiabatic entropy $s_{\rm ad}$ derived from 3D simulations to calibrate $\alpha_{\rm MLT}$ throughout the evolution calculation. In this section, we compare the three methods by summarizing their reasoning and methodology, and pointing out their respective advantages and limitations.  
  
  The 1D-3D coupling method has two outstanding advantages. By replacing near-surface layers of the 1D stellar structure model with the mean 3D model atmosphere, the model-predicted theoretical frequencies agree better with asteroseismic observations. Placing the matching point in the near-adiabatic region also makes the evolution tracks insensitive to the choice of $\alpha_{\rm MLT}$, effectively eliminating this important adjustable parameter. Another accompanying benefit is that consistent opacity between 1D and 3D models is not essential for this method, as the flux of radiative diffusion is negligible compared to the total energy flux at the matching location.
  
  On the other hand, since the matching point is located in the hydrogen ionization zone ($T_{\rm m} > 10000$ K), where thermodynamics quantities are relatively more sensitive to the adopted EOS and the hydrogen mass fraction, using identical EOS between the 1D and 3D models is advantageous. It is worth noting that element diffusion alters hydrogen, helium and metal mass fractions at the outermost shell of the 1D model, whereas the relationship between $X$ and $Z$ is fixed by the chemical composition employed in the 3D simulation. Although our metallicity interpolation ensures that $Z/X$ is identical between the atmosphere and interior model, individual H, He and metal mass fractions are not necessarily consistent between the two parts. This is a common problem when using tabulated model atmospheres as outer boundary conditions for stellar evolution. Nevertheless, we note that \citet{2021MNRAS.504.3128S} and \citet{2024A&A...687A.146M} have taken into account the change of mean molecular weight in their entropy calibration, as discussed below.
  Finally, the 1D-3D coupling method requires not only the mean temperature and pressure but also partial derivatives at the matching point, making this method more susceptible to the interpolation scheme and the density of the 3D grid compared to other approaches. 
  
  \citet{2014MNRAS.442..805T,2014MNRAS.445.4366T} conceived and developed a different approach to improve stellar evolution calculation using information from 3D simulations. Being aware that convection is described in a fundamentally different way in 1D models and 3D hydrodynamics simulations, \citet{2014MNRAS.442..805T} extracted the radiative $T(\tau)$ relation that excludes the contribution from convection from their 3D grid. The $\alpha_{\rm MLT}$ calibration is carried out using a stellar envelope code with identical microphysics. At each $T_{\rm eff}$ and $\log g$, the appropriate $\alpha_{\rm MLT}$ is determined by requiring the temperature of the envelope and the 3D model to agree at the matching point. The radiative $T(\tau)$ relation provides atmosphere boundary conditions for stellar evolution calculations and meanwhile affects the temperature stratification near the photosphere, and should be applied in conjunction with the calibrated $\alpha_{\rm MLT}$. Similar to standard stellar evolution, the outer boundary is located at the photosphere (but not $\tau = 2/3$ as in the case of the gray atmosphere), above which atmosphere integration is performed based on the $T(\tau)$ relation. This method relies on tabulated $T(\tau)$ relations and $\alpha_{\rm MLT}$ values, making it less affected by the choice of interpolation method and less prone to numerical issues compared to the 1D-3D coupling approach.
  However, when utilizing the 3D-calibrated $T(\tau)$ relation and $\alpha_{\rm MLT}$ in stellar evolution calculation, it is preferred to use identical EOS and opacity as the 3D grid. If the stellar evolution code adopts different microphysics, an additional scaling factor to the calibrated mixing-length parameter is necessary to compensate for this inconsistency (see \citealt{2018MNRAS.478.5650M} Eq.~2). As investigated in \citet{2018MNRAS.478.5650M}, implementing the 3D-calibrated $T(\tau)$ relation and $\alpha_{\rm MLT}$ shifts the evolutionary track to slightly hotter $T_{\rm eff}$ in the RGB relative to the gray atmosphere. The improved structural model, however, has minimal effect on the predicted p-mode frequencies \citep[Fig.~7]{2018MNRAS.478.5650M}, because the near-surface layers are still described by the MLT.
  
  The entropy calibration method proposed and implemented by \citet{2016ApJ...822L..17T}, \citet{2018ApJ...869..135S,2021MNRAS.504.3128S} and \citet{2024A&A...687A.146M} is essentially an on-the-fly calibration of $\alpha_{\rm MLT}$ during stellar evolution. At every evolution time step, the mixing-length parameter is adjusted such that entropy at the bottom convective boundary equals the entropy of the inflow at the bottom boundary of the simulation (referred to as asymptotic entropy in \citealt{2015A&A...573A..89M}) at that $T_{\rm eff}$, $\log g$ and metallicity. The asymptotic entropy was expressed as a function of global stellar parameters for easy implementation. As $\alpha_{\rm MLT}$ is fixed by the entropy calibration, it is no longer a free parameter within this approach.
  Possible inconsistencies in chemical composition between the 1D and 3D models are mended by introducing an extra factor based on the mean molecular weight \citep{2021MNRAS.504.3128S,2024A&A...687A.146M}. Moreover, \citet{2018ApJ...869..135S} and \citet{2024A&A...687A.146M} included a constant correction factor to account for different entropy ``zero points'' defined in EOSs. \citet{2024A&A...687A.146M} showed that the entropy correction factor does not vary significantly across the parameter space, suggesting different EOSs between 1D and 3D models are acceptable within this approach provided that the correction factor is carefully determined. Finally, like the calibrated $T(\tau)$ and $\alpha_{\rm MLT}$ method, the entropy calibration is anchored to surface parameters such as $T_{\rm eff}$ and the adiabatic entropy, hence will not lead to notable improvement on theoretical oscillation frequencies. 
  
  To sum up, the major advantage of the 1D-3D coupling approach compared to other methods is that near-surface layers are represented by the mean 3D model which is more realistic, making our method suitable for modeling stars with asteroseismic data. On the other hand, the calibrated $T(\tau)$ relation and $\alpha_{\rm MLT}$ as well as the entropy calibration method, especially the latter, is less influenced by numerical challenges. It will be worthwhile to implement all three approaches to the same stellar evolution code for quantitative comparisons of evolutionary tracks and interior structures predicted by each method.

\section{Conclusions}

  As a continuation of papers I and II \citep{2018MNRAS.481L..35J,2020MNRAS.491.1160M}, we refined and further developed the method of coupling 1D stellar evolution with 3D model atmosphere on-the-fly by implementing the interpolation of mean 3D models across metallicities in \garstec{}. This effort allows us to utilize the majority of \stagger{}-grid models and extend the capability of the 1D-3D coupling method to stars with non-solar metallicity. We demonstrate quantitatively that by placing the outer boundary of the stellar interior model well beneath the stellar surface in the near-adiabatic layer, stellar evolution tracks computed with the 1D-3D coupling method become far less sensitive to $\alpha_{\rm MLT}$ compared to the standard approach. The consequence is that $\alpha_{\rm MLT}$ can be regarded as a constant determined by solar calibration.
  
  The 1D-3D coupling method was validated against stars in detached eclipsing binary systems, where accurate stellar mass and radius could be measured from binary orbits. It is clear from Sect.~\ref{sec:validation} that the method can be applied for realistic evolutionary modeling of stars with different metallicities. 
  With $\alpha_{\rm MLT}$ fixed and few input parameters to adjust, our method is capable of generating stellar models that reproduce the basic properties for both components of AI Phe, whose fundamental parameters are extremely well constrained. Tests on oscillating \textit{Kepler} giants in binary systems highlight the strength of the 1D-3D coupling method. For both giants, their radial and mixed mode frequencies are fitted by our coupled models at a satisfactory level, without the need for artificial surface corrections. Although the surface term is still not fully eliminated, our method of coupling 1D and 3D models during stellar evolution demonstrates great potential for modeling stars with precise asteroseismic data.
  The best models of two \textit{Kepler} giants have slightly higher masses than inferred from observations. As an overestimated mass leads to a younger stellar age, this systematic offset deserves further investigation in future studies.
  
  Meanwhile, bends (or kinks) observed in the evolutionary tracks indicate room for improvement in the 1D-3D coupling method. An optimum numerical approach for interpolating mean 3D models and evaluating partial derivatives at the outer boundary is necessary. Of course, the reliability of the method and accuracy of the interpolation would benefit from increasing the density of the grid in the $(T_{\rm eff}, \log g)$ plane, and supplying missing 3D models at the edges of the grid (e.g.~$T_{\rm eff} = 5000$ K, $\log g = 2$) would allow metal-poor tracks to reach higher up the RGB. Nevertheless, preliminary investigations with a denser grid indicate that this would typically result in changes less than about 20 K in the effective temperature.
  Another open problem in metallicity interpolation is that the method is based on the mass fraction ratio $Z/X$. As a result, individual hydrogen, helium, and metal mass fractions may differ between the 1D and 3D models when element diffusion is taken into account.
  We will address these issues in future work to fully exploit the potential of the 1D-3D coupling method for modeling oscillating stars and determining the ages of star clusters.

  The fact that our 1D-3D coupling models remove the need for $\alpha_{\rm MLT}$ in calculating thermal structure near the surface introduces new opportunities for stellar modeling. This is particularly relevant for red giants, where $\alpha_{\rm MLT}$ exhibits strong degeneracies with $Y_{\rm init}$, and as a result affecting stellar age estimates by up to 30\%. This new approach could lead to more accurate stellar age estimates as well as potentially tighter constraints on helium abundance --- both are crucial for understanding the history of Galactic chemical enrichment. 
  Additionally, since stellar fundamental parameters such as radius and $T_{\rm eff}$ are insensitive to $\alpha_{\rm MLT}$, the frequency of maximum oscillation power estimated from the $\nu_{\rm max} \propto g/\sqrt{T_{\rm eff}}$ scaling relation is less affected by modeling uncertainties, permitting more meaningful comparisons against observed $\nu_{\rm max}$. This provides a means to test the validity of the $\nu_{\rm max}$ scaling relation in red giants. Such a test would be very valuable for metal-poor stars, where the $\nu_{\rm max}$ scaling relation remains questionable to use but is highly needed.

\section*{Acknowledgements}
  The authors are grateful to Jeppe Thomsen for providing observational data of KIC 10001167, and to Anish Amarsi and Mia Lundkvist for reading and commenting on this manuscript. We thank also Meridith Joyce, Victor Aguirre B{\o}rsen-Koch, Regner Trampedach and Teresa Braun for valuable comments and fruitful discussions. Y.Z.~thanks the hospitality of Max Planck Institute for Astrophysics during his visit.
  Y.Z.~gratefully acknowledges support from the Elaine P.~Snowden Fellowship. A.C.S.J.~is supported by the Eric and Wendy Schmidt AI in Science Postdoctoral Fellowship, a Schmidt Sciences program. CL acknowledges funds from the UKRI Future Leaders Fellowship (MR/S035214/1). CL also acknowledged funds from the European Research Council (ERC) under the European Union's Horizon 2020 research and innovation programme (Grant agreement No. 852977). O.K.~acknowledges support by the Swedish Research Council (grant agreement 2023-03667) and the Swedish National Space Agency. This work was supported by a research grant (42101) from VILLUM FONDEN. Funding for the Stellar Astrophysics Centre was provided by The Danish National Research Foundation (Grant DNRF106). The numerical results presented in this work were partly obtained at the Centre for Scientific Computing, Aarhus \url{https://phys.au.dk/forskning/faciliteter/cscaa/}. This research was supported by computational resources provided by the Australian Government through the National Computational Infrastructure (NCI) under the National Computational Merit Allocation Scheme and the ANU Merit Allocation Scheme (project y89).

\section*{Data Availability}

The data underlying this article will be shared on reasonable request to the corresponding author.



\bibliographystyle{mnras}
\bibliography{References}




\appendix

\section{Interpolation errors for mean 3D models} \label{app:itp-error}

\begin{table*} \label{tb:itp-test}
\centering
\caption{Effective temperature and surface gravity of \mean{\rm 3D} models used in the interpolation test. Except for the solar model, model names are constructed from the effective temperature (labeled as ``\texttt{t}''), surface gravity (``\texttt{g}''), and ``\texttt{m00}'' stands for solar metallicity. Effective temperatures listed here are the mean value averaged from all simulation snapshots. We note that models \texttt{t62g43m00} and \texttt{t47g27m00} are not included in the \stagger{}-grid (Fig.~\ref{fig:grid}).
\label{tb:Teff-logg}}
{\begin{tabular*}{0.7\textwidth}{@{\extracolsep{\fill}}lcccccc}
\toprule[2pt]
  Model name 		& \texttt{solar}		& \texttt{t62g43m00}		& \texttt{t52g42m00}		& \texttt{t47g47m00}		& \texttt{t47g32m00}		& \texttt{t47g27m00}
  \\
\midrule[1pt]
  $T_{\rm eff}$ (K) & 5776	& 6231	& 5242	& 4743	& 4766	& 4695
  \\
  $\log g$ (cgs)    & 4.44 	& 4.319 	& 4.25	& 4.75	& 3.25	& 2.67
  \\
\bottomrule[2pt]
\end{tabular*}}
\end{table*}

\begin{figure*}
\includegraphics[width=0.49\textwidth]{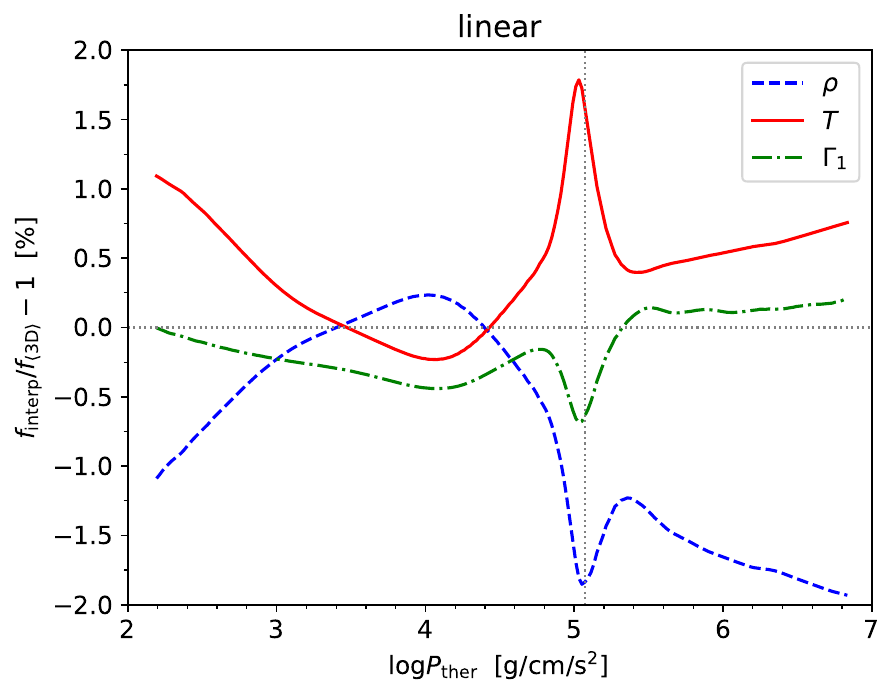}
\includegraphics[width=0.49\textwidth]{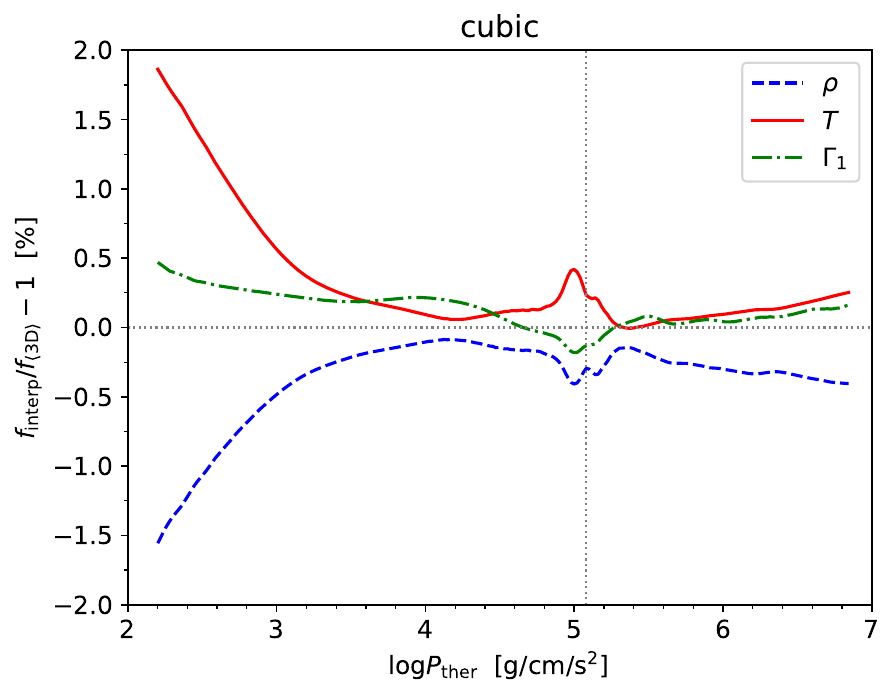}
\caption{Interpolation error of density, temperature, and the first adiabatic exponent $\Gamma_1$ for the mean 3D solar model. Error from linear (cubic) interpolation at different thermal pressure is shown in the \textit{left} (\textit{right}) panel. The vertical dotted line marks the location of density inflection defined as the local minimum of $\partial\ln\rho / \partial\ln P_{\rm ther}$. Below the density inflection, the error in temperature ranges from about 0.5\% to over 0.7\% for linear interpolation and less than 0.25\% for cubic interpolation.
\label{fig:itp-diff-solar}}
\end{figure*}

\begin{figure*}
\includegraphics[width=0.49\textwidth]{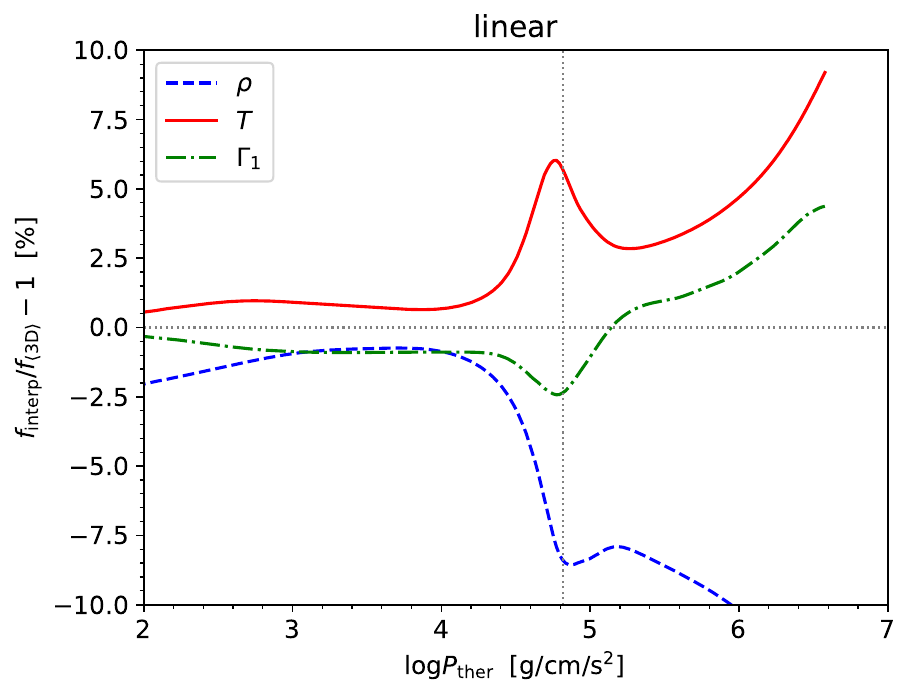}
\includegraphics[width=0.49\textwidth]{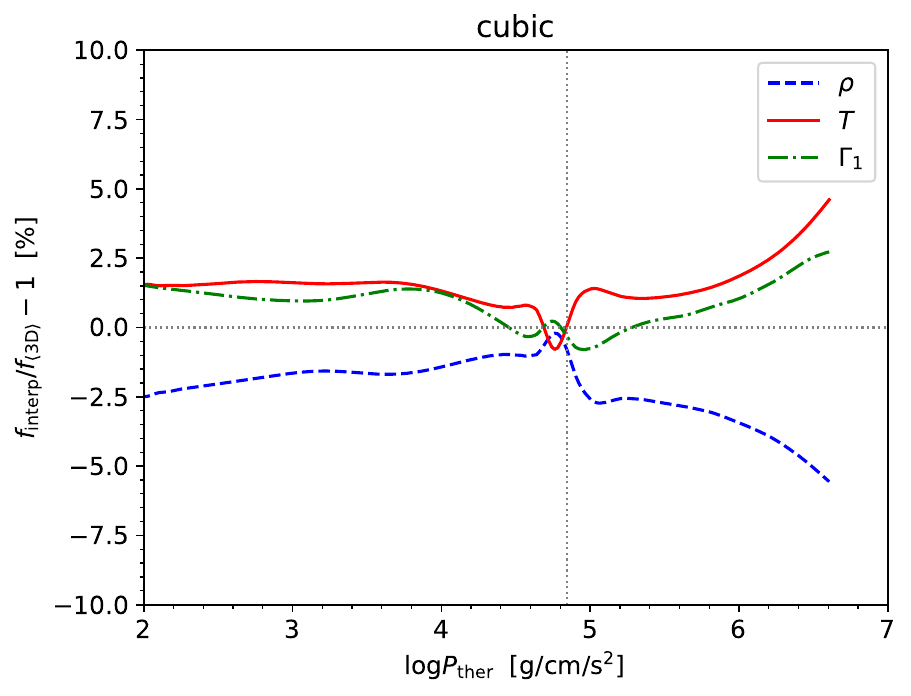}
\caption{Interpolation error for the mean structure of model \texttt{t62g43m00} (see Table \ref{tb:Teff-logg}). The error of linear interpolation is larger than the cubic method around and below the density inflection. In deeper layers, relative error from linear interpolation reaches up to 9\% for temperature, while the cubic method results in less than 5\% error.
\label{fig:itp-diff-t62g43m00}}
\end{figure*}

\begin{figure*}
\includegraphics[width=0.49\textwidth]{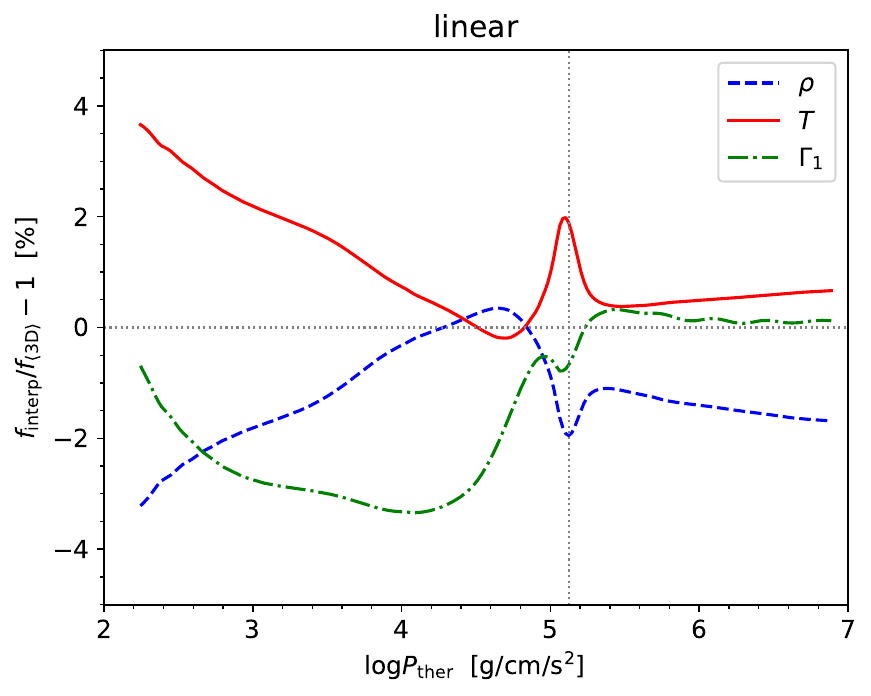}
\includegraphics[width=0.49\textwidth]{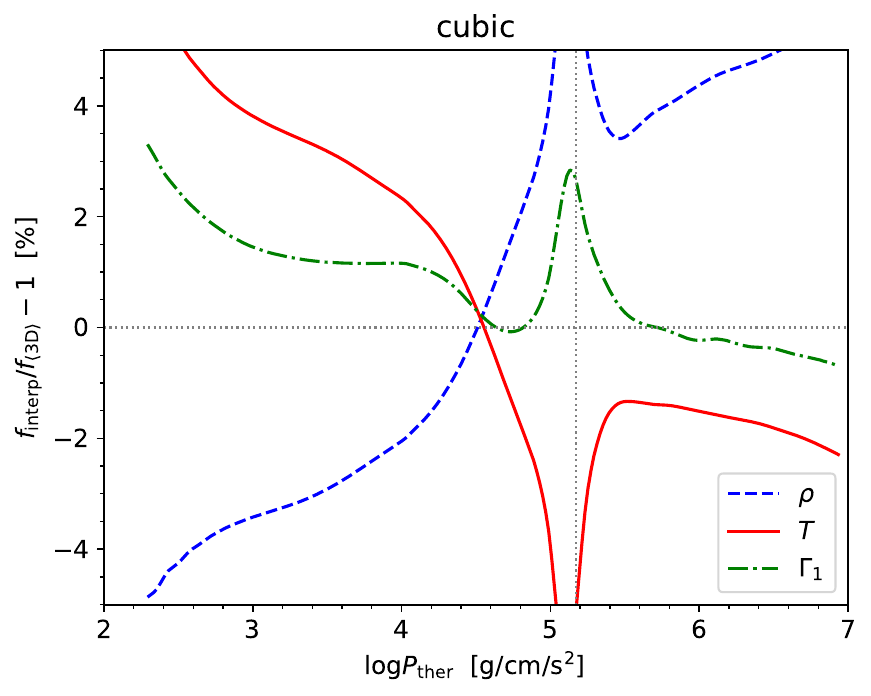}
\caption{Interpolation error for the mean structure of model \texttt{t52g42m00}. Below the density inflection, the error of cubic interpolation is larger than the linear method: Temperature profile of the \mean{\rm 3D} model differs from the linear interpolated values by less than 0.75\%. The relative differences increase to nearly 2\% for the cubic method.
\label{fig:itp-diff-t52g42m00}}
\end{figure*}

\begin{figure*}
\includegraphics[width=0.49\textwidth]{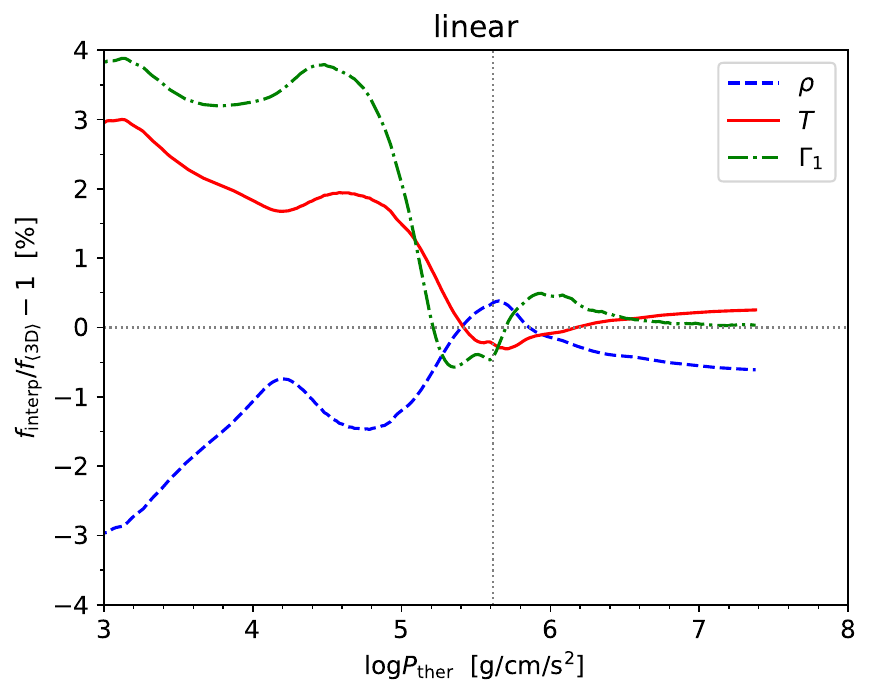}
\includegraphics[width=0.49\textwidth]{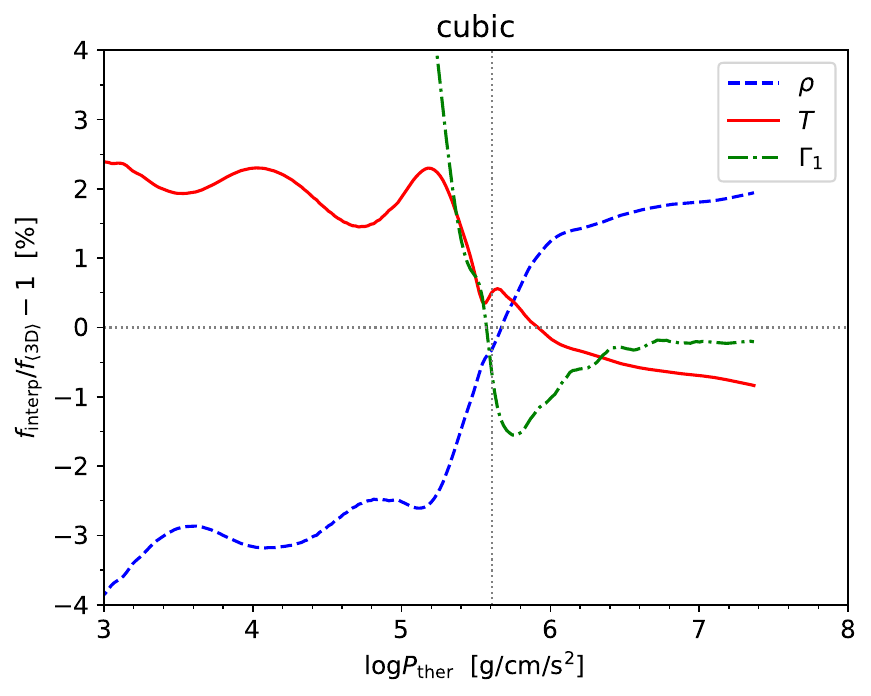}
\caption{Interpolation error for the mean structure of model \texttt{t47g47m00}.
\label{fig:itp-diff-t47g47m00}}
\end{figure*}

\begin{figure*}
\includegraphics[width=0.49\textwidth]{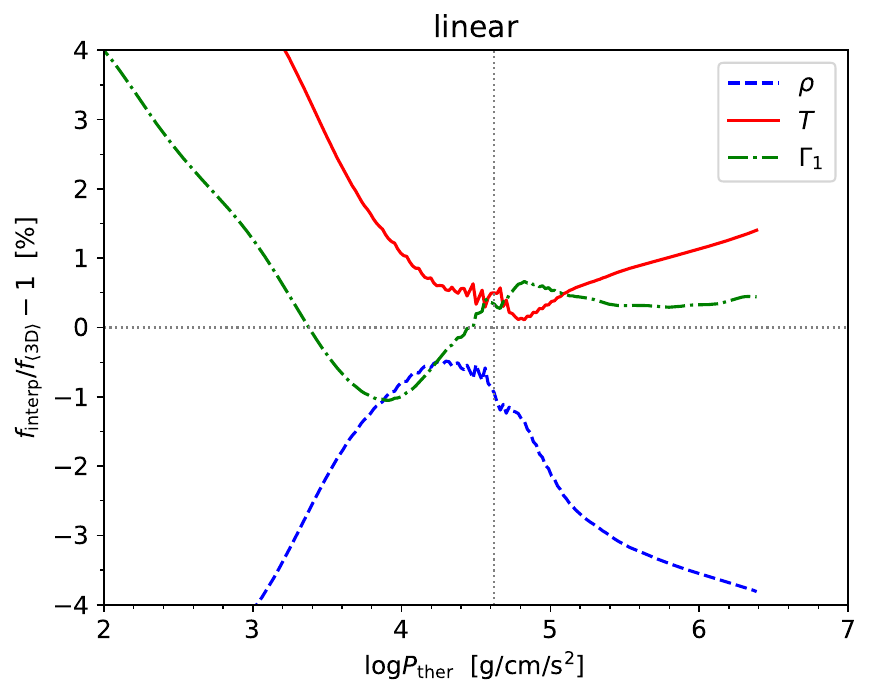}
\includegraphics[width=0.49\textwidth]{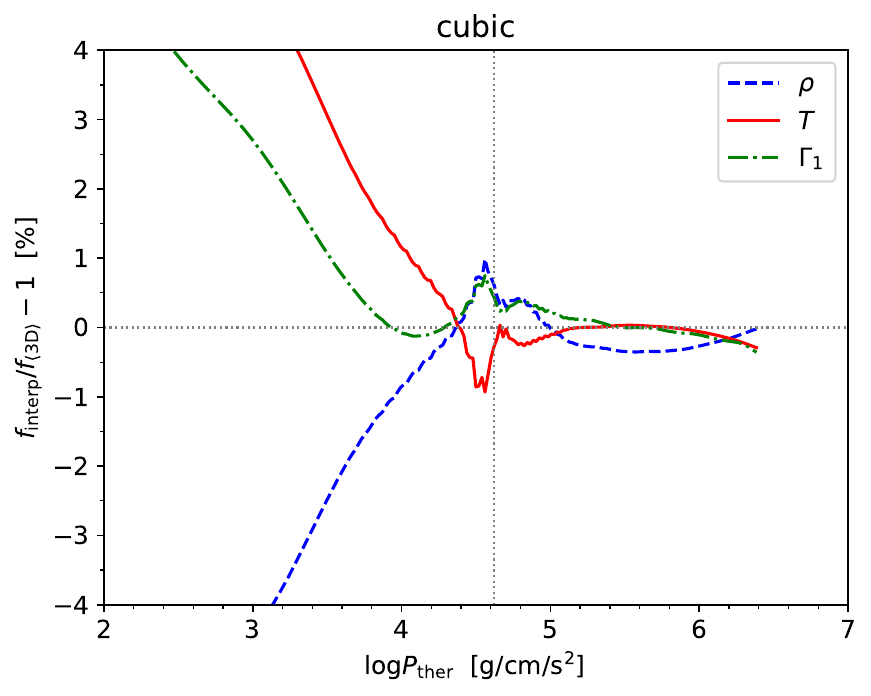}
\caption{Interpolation error for the mean structure of model \texttt{t47g32m00}. Below the density inflection, the error in temperature ranges from less than 0.25\% to about 1.5\% for linear interpolation and less than 0.25\% for the cubic method.
\label{fig:itp-diff-t47g32m00}}
\end{figure*}

\begin{figure*}
\includegraphics[width=0.49\textwidth]{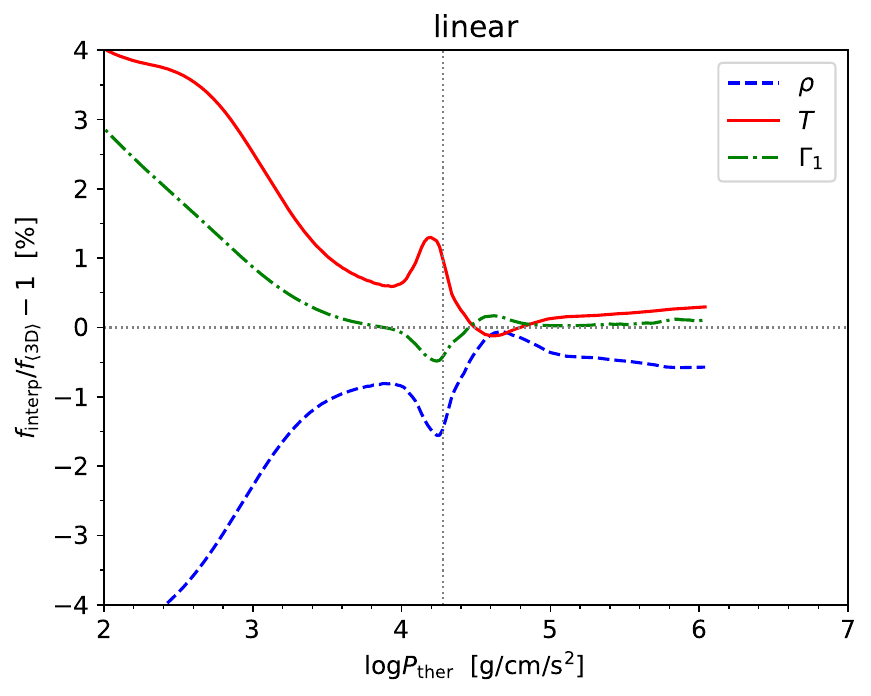}
\includegraphics[width=0.49\textwidth]{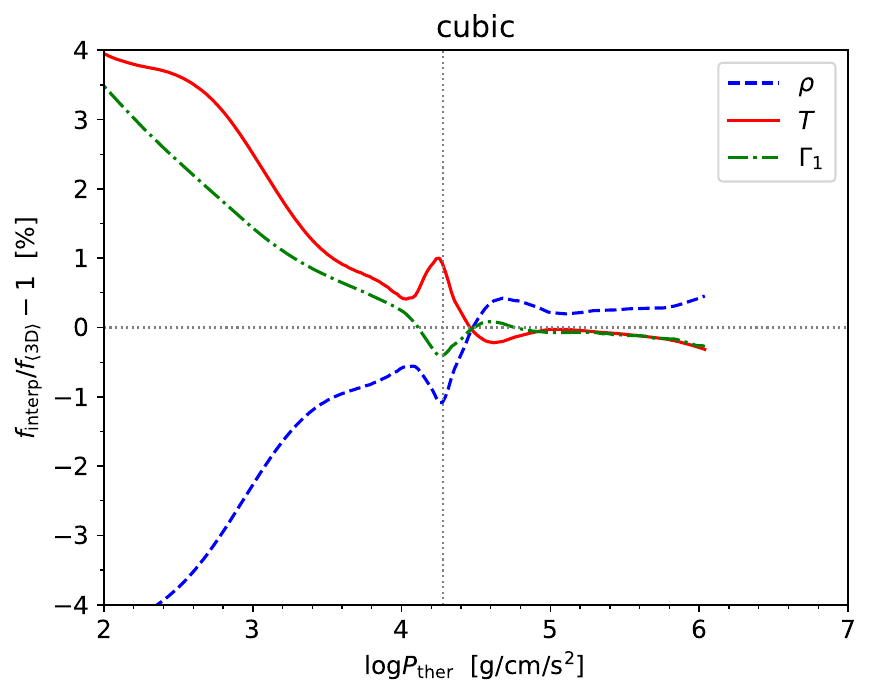}
\caption{Interpolation error for the mean structure of model \texttt{t47g27m00}.
\label{fig:itp-diff-t47g27m00}}
\end{figure*}

  We quantify the errors of our interpolation scheme introduced in Sect.~\ref{sec:3D-model} across the HR diagram by taking one \mean{\rm 3D} model out from the grid before constructing an interpolated mean stratification at the corresponding surface temperature and gravity based on the remaining models, or by comparing our interpolation results with an additional 3D model not included in the grid. As shown in Table~A1, multiple tests are conducted for the solar parameter, models correspond to warm (F-type) and cool (K-type) dwarfs as well as red giants. Given the 3D grid adopted in this work, the optimal interpolation method (linear or cubic) depends on the stellar parameters.


\bsp	
\label{lastpage}
\end{document}